\documentclass[preprint,aps]{revtex4}
\usepackage{graphicx}
\usepackage{dcolumn}
\usepackage{bm}

\begin{document}

\title{Analytical  Solutions for the  Nonlinear Longitudinal Drift Compression (Expansion) of Intense Charged Particle Beams}
\author{Edward A. Startsev}
\author{Ronald C. Davidson}
\affiliation{Plasma Physics Laboratory\\ Princeton University\\
Princeton, NJ 08543}
\date{\today}
\begin{abstract}
To achieve  high focal spot intensities in heavy ion fusion, the
ion beam must be compressed longitudinally by factors of ten to
one hundred before it is focused onto the target. The longitudinal
compression is achieved by imposing an initial velocity profile
tilt on the drifting beam. In this paper, the problem of
longitudinal drift compression of intense charged particle beams
is solved analytically for the two important cases corresponding
to a cold beam, and a pressure-dominated beam, using a
one-dimensional warm-fluid model describing the longitudinal beam
dynamics.
\end{abstract} \keywords{}

\maketitle 

\section{Introduction}

High energy ion accelerators, transport systems and storage rings
\cite{davqin01,rei94,law88,chao93,edw93} have a wide range of
applications ranging from basic research in high energy and
nuclear physics, to applications such as heavy ion fusion,
spallation neutron sources, and nuclear waste transmutation.
 Of particular importance at
the high beam currents and charge densities of interest for heavy
ion fusion are the effects of the intense self-fields produced by
the beam space charge and current on determining detailed
equilibrium, stability, and transport properties. In general, a
complete description of collective processes in intense charged
particle beams is provided by the nonlinear Vlasov-Maxwell
equations \cite{davqin01} for the self-consistent evolution of the
beam distribution function, $f_b(x,{\bf p},{\bf t})$, and the
self-generated electric and magnetic fields, ${\bf E}({\bf x},{\bf
t})$ and $ {\bf B}({\bf x},{\bf t})$. While considerable progress
has been made in analytical and numerical simulation studies of
intense beam propagation \cite{kap59,glu71,wang82,hof83,
str84,hof87,str92,bro95,davqin99,glu95,davchen98,
chen97,chen94,davlee98,davprl98,dav98,sto99,lee97,qian97,tze02,
dav01,davchan99,dav02,har59,stardav02,start02,start03,start04,kish00,hab99,
frie92,wei59,startdav03,dav72,lee73,kap74,hon00,neil65,lee81,lee92,
dav03,neu92,mac01,dav99,davsto99,davqin00,qindav00,davuhm01,qindav03,
wang03,qinstart03,qin03,lee78,lau78,ros60,uhm80,uhm81,fern95,uhm03,
uhm01,joy83,uhmdav80,kag01,kag02,rose99,wel02}, the effects of
finite geometry and intense self-fields often make it difficult to
obtain detailed predictions of beam equilibrium, stability, and
transport properties based on the Vlasov-Maxwell equations. To
overcome this complexity, considerable theoretical progress has
also been made in the development and application of
one-dimensional Vlasov-Maxwell models
\cite{gf1,gf2,gf3,gf4,gf5,gf6,gf7,gf8} to describe the
longitudinal beam dynamics for a long coasting beam, with
applications ranging from plasma echo excitations, to the
investigation of coherent soliton structures, both compressional
and rarefactive (hole-like).  Such one-dimensional Vlasov
descriptions rely on using a geometric-factor ($g$-factor) model
\cite{gf8,gf9,gf10,gf11,gf12,gf13,gf14} to incorporate the average
effects of the transverse beam geometry and the surrounding wall
structure. Despite the many  successful applications of such
one-dimensional Vlasov models to describe the longitudinal
dynamics of  long costing beams, even a one-dimensional Vlasov
description is often too complicated to be analyzed in detail,
except in some simple limiting cases. The next level of
description of  the nonlinear beam dynamics is provided by the
macroscopic fluid equations, which correspond to the first three
momentum moments of the Vlasov equation, with  some particular
closure scheme which relates the  higher moments to the first
three \cite{dav1,dav2,dav3}. The usual closure assumption  for
collisionless plasma is provided by the adiabatic equation of
state ($ds/dt=0$, where $s$ is the entropy per the unit volume),
which expresses the thermal pressure as a function of the density.
 In one dimension, this  is given by
$p\lambda^{-3}=const.$ \cite{dav1,dav2,dav3}, where $\lambda$ is
the line density of beam particles and $p$ is the line pressure.
Such a one-dimensional fluid model, combined with the adiabatic
equation of state and a g-factor model for the average electric
field, fits in the class of  one-dimensional fluid problems which
can be solved exactly \cite{lan1} using the formalism described in
Sec. II.

\begin{figure}[htb]
\centering
\includegraphics*[width=90mm]{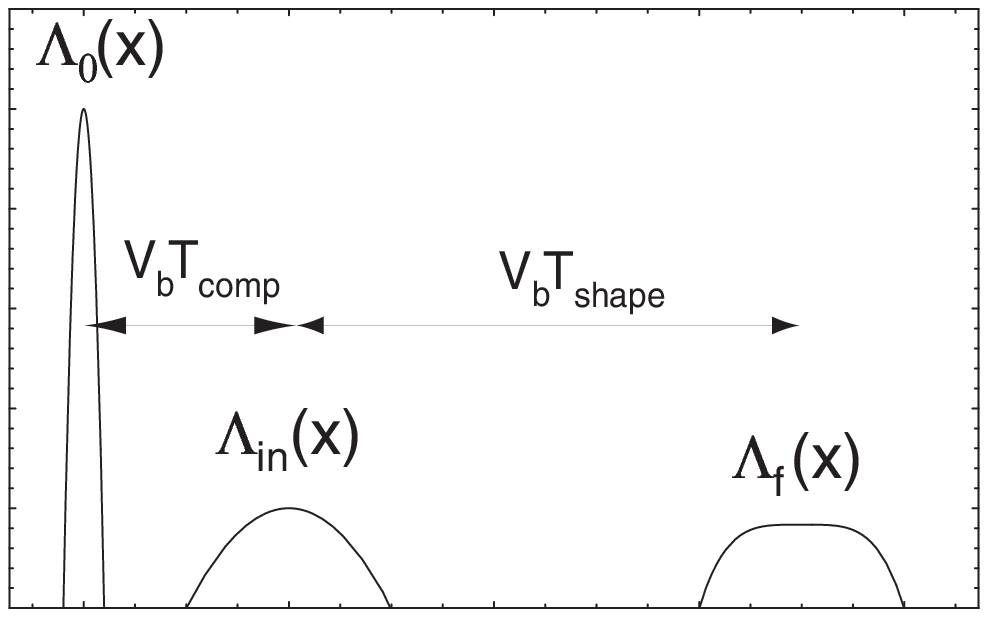}
\vspace{400pt} \caption{}\label{fig1}
\end{figure}

 In presently envisioned configurations  for heavy ion
fusion, multiple, high-current, heavy ion  beams are focused to a
small spot size onto the target capsule. To achieve a
high-intensity beam focused onto the target, the beams are first
accelerated and then compressed longitudinally. One of the
possible ways to compress the beam longitudinally is to use a the
drift compression scheme illustrated in Fig.\ \ref{fig1}
\cite{hong1,hong2,hong3,hong4,hong5,hong6,hong7,hong8,hong9,hong10}.
The scheme consists of two parts. In the first stage, an initial
tilt in the longitudinal velocity profile $-V_f(x)$ is imposed on
the long charge bunch  with some particular line density profile
$\Lambda_f(x)$. After time $T_{shape}$, the beam line density
evolves to a  profile $\Lambda_{in}(x)$, with  velocity profile
$-V_{in}(x)$. At this time, an additional velocity tilt
$V_{in}(x)-V_{comp}(x)$ is imposed on the beam, and the beam is
left with  line density $\Lambda_{in}(x)$ and velocity tilt
$-V_{comp}(x)$. We refer to this stage as the {\it beam shaping}
stage. This stage requires beam manipulation (imposing the
velocity tilt) and is done when the charge bunch is very long. At
this stage, the longitudinal pressure and electric field are
negligible, and the beam dynamics is governed by free convection.
During this stage the beam may or may not be compressed. The
purpose of this stage is to shape the beam line density profile
into  a certain intermediate line density profile
$\Lambda_{in}(x)$ with  velocity profile $-V_{comp}(x)$, such that
during the next stage,  which we refer to as the  {\it drift
compression} stage, after further axial drift during the time
interval  $T_{comp}$ (see Fig.\ \ref{fig1}), the beam is
compressed longitudinally until the space-charge force or the
internal thermal pressure stops the longitudinal compression of
the charge bunch
\cite{hong1,hong2,hong3,hong4,hong5,hong6,hong7,hong8,hong9,hong10}.
At the point of maximum compression, the velocity tilt profile is
completely removed and the beam is left with desired line density
$\Lambda_0(x)$. The final focus magnets then focus the beam onto
the target, and the beam heats and compresses the target fuel.
Stated this way, the longitudinal drift compression problem in the
beam frame is equivalent to the time-reversed problem of the beam
expanding into vacuum with zero initial velocity profile, and the
specified initial line density profile $\Lambda_0(x)$ which is
desired at the time of maximum compression before final focusing.
In this paper we employ a one-dimensional warm-fluid model with
adiabatic equation of state to study this problem analytically for
arbitrary final (maximally compressed) line density profiles
$\Lambda_0(x)$.

We consider here the two separate cases corresponding to a  cold
beam, and a  pressure-dominated beam. In the case of a cold beam,
the internal thermal  pressure is negligible, and the dynamics of
the drift compression is governed by the self-generated electric
field. In the case of a pressure-dominated beam, the
self-generated electric field is negligible, and the beam
compresses under the influence of the thermal pressure and the
initial velocity tilt $-V_{comp}(x)$. One of the compression
scenarios considered for heavy ion fusion is   {\it neutralized}
drift compression, where the beam propagates through a
charge-neutralizing background plasma as it compresses
longitudinally. In such a scenario, the beam is compressed only
against the internal pressure. For simplicity, the present
analysis is carried out in the beam frame where the particle
motions are nonrelativistic. The final results can be  then
Lorentz transformed back to the laboratory frame, moving with
axial velocity $-V_b=-\beta _bc$ relative to the average motion of
the particles in the beam frame.

 This paper is organized
as follows. In Sec. II, we briefly describe the one-dimensional
warm-fluid model equations and the formalism used for solving them
analytically. In Secs. III and IV, the general solution for the
expansion problem (the inverse to the {\it drift compression
stage} problem) is obtained analytically for the two cases
corresponding to a cold beam, and a pressure-dominated beam. In
Sec. V the general solution to drift compression stage obtained in
Secs. III and IV is illustrated by several examples. In Sec. VI we
briefly discuss the {\it beam shaping} stage.

\section{Theoretical Model}
In the present analysis, we employ a one-dimensional warm-fluid
model \cite{dav1,dav2,dav3,hong2} to describe the longitudinal
nonlinear beam dynamics with average  electric field given by the
g-factor model with $e_bE_z=-e_b^2g\partial \lambda/\partial x$
\cite{gf8,gf9,gf10,gf11,gf12,gf13,gf14}. For example, for a
space-charge-dominated beam with flat-top density profile in the
transverse plane, $g\simeq 2\ln(r_w/r_b)$ \cite{gf14}. Here,
$\lambda(x,t)$ is the line density, $e_b$ is the charge of a beam
particle, $r_w$ is the conducting wall radius, and $r_b$ is the
beam radius. Generally, the beam radius, and therefore the
$g$-factor,  are functions of the line density and the external
transverse focusing, and can change during the beam compression.
In most of the drift compression scenarios it is preferable to
maintain the beam radius (and therefore the g-factor) constant
during the beam compression by adjusting  the external transverse
focusing \cite{hong1,hong2,hong3}. Therefore, in the present
analysis, we assume that the g-factor is a constant.

The macroscopic fluid equations for the line density
$\lambda(x,t)$, the average longitudinal beam velocity $v(x,t)$,
and the longitudinal line pressure $p(x,t)$ are given by
\cite{dav1,dav2,dav3,hong2}
\begin{eqnarray}
&&{\partial \lambda \over \partial t}+{\partial \over \partial x}
(\lambda v)=0,\label{eq1}\\
&&{\partial v \over \partial t}+v{\partial \over
\partial x}v =-{e_b^2 g\over m_b}\ {\partial \lambda \over
\partial x}-{1\over m_b \lambda}\ {\partial p \over \partial
x}=-{\partial w \over \partial x},\label{eq2}
\end{eqnarray}
where  $p=(p_0/\lambda_0^3)\lambda^3$ for a triple-adiabatic
equation of state. Here we have introduced the effective potential
$w$ defined by
\begin{equation}
w=c_g^2{\lambda\over\lambda_0}+{c_p^2\over 2}{\lambda^2\over
\lambda_0^2},\label{eq2a}
\end{equation}
where $c_g^2=e_b^2g\lambda_0/m_b$ and $c_p^2=3 p_0/2 m_b\lambda_0$
are constants with dimensions of $(\text{speed})^2$, $m_b$ is the
mass of a beam particle,  and $\lambda_0$ and $p_0$ are constants
with the dimensions of  line density and line pressure,
respectively.

For future application in Secs. III-VI, in the remainder of this
section we summarize well-established theoretical technique
developed in fluid mechanics \cite{lan1} that can be used to solve
the nonlinear fluid equations (\ref{eq1}) and (\ref{eq2}). By
introducing the velocity potential $\phi$, where $v=\partial
\phi/\partial x$, we can rewrite Eq.\ (\ref{eq2}) as
\begin{equation}
{\partial \phi \over \partial t}+{v^2\over 2} +w =0.\label{eq3}
\end{equation}
The full differential of $\phi$ then becomes
\begin{equation}
d\phi={\partial \phi \over \partial x}dx+{\partial \phi \over
\partial t}dt=v dx -\left({v^2\over 2} +w\right)dt.
\end{equation}\label{eq4}Next, following  Landau and Lifshitz \cite{lan1}, we introduce the
Legendre transform
\begin{equation}
d\phi=d(x v)-x dv -d \left[t\left({v^2\over 2}
+w\right)\right]+td\left({v^2\over 2} +w \right).\label{eq5}
\end{equation}
Introducing $\chi=\phi -xv -t(w+v^2/2)$,  Eq.\ (\ref{eq5}) can be
expressed as
\begin{equation}
d\chi=-x dv + td\left({v^2\over 2} +w \right)=t dw +(v t-x) dv
.\label{eq6}
\end{equation}
It follows from Eq.\ (\ref{eq6}) that  $\chi$  can be considered
as a function of the new independent variables $(v,w)$, and that
\begin{eqnarray}
t={\partial \chi \over \partial w},\  x-vt=-{\partial \chi \over
\partial v}.\label{eq7}
\end{eqnarray}
Therefore, if the function  $\chi$ is known as a  function of its
arguments $(v,w)$, then Eq.\ (\ref{eq7}) gives $(v,w)$ as implicit
functions of $(x,t)$.

To obtain the equation for $\chi$ we rewrite Eq.\ (\ref{eq1}) as
\begin{equation}
{\partial(\lambda,x)\over \partial
(t,x)}+v{\partial(t,\lambda)\over
\partial (t,x)}+\lambda {\partial(t,v)\over \partial (t,x)}=0,
\label{eq8}
\end{equation}
where
\begin{equation}
{\partial(a,b)\over \partial (x,y)}\equiv{\partial a \over
\partial x}{\partial b \over \partial y}-{\partial a \over
\partial y}{\partial b \over \partial x}.\label{eq9}
\end{equation}
Assuming that $v$ is not a definite function of $w$ [$v\neq
v(w)$], we  multiply Eq.\ (\ref{eq8}) by $\partial(t,x)/
\partial (w,v)$ and use the multiplication property for
determinants. This gives
\begin{equation}
{\partial(\lambda,x)\over \partial
(w,v)}+v{\partial(t,\lambda)\over
\partial (w,v)}+\lambda {\partial(t,v)\over \partial (w,v)}=0.
\label{eq10}
\end{equation}
The case when $v$ is a definite function of $w$ in some region of
the $(x,t)$ plane corresponds to {\it a simple wave} and will be
considered later. Because $\lambda=\lambda(w)$, Eq.\ (\ref{eq10})
reduces to
\begin{equation}
{d \lambda \over dw}{\partial x \over \partial v}-v{d \lambda
\over dw}{\partial t \over \partial v}+\lambda {\partial t \over
\partial w}=0.\label{eq11}
\end{equation}
Substituting Eq.\ (\ref{eq7}) into Eq.\ (\ref{eq11}), we obtain
the equation for  $\chi$ \cite{lan1}
\begin{equation}
{1\over \lambda} {d \lambda \over dw}\left( {\partial \chi\over
\partial w}-{\partial ^2 \chi\over\partial
v^2}\right)+{\partial ^2 \chi\over\partial w^2}=0.\label{eq12}
\end{equation}
Note that Eq.\ (\ref{eq12}) is a linear partial differential
equation for  the function $\chi(v,w)$. By introducing the
effective sound speed defined by $c^2=\lambda dw/d\lambda$, we can
rewrite  Eq.\ (\ref{eq12}) as
\begin{equation}
{\partial \chi\over \partial w}-{\partial ^2 \chi\over\partial
v^2}+c^2{\partial ^2 \chi\over\partial w^2}=0,\label{eq13}
\end{equation}
where $c^2$ is to be regarded as a function of $w$. Equation\
(\ref{eq13})
 together with Eq.\ (\ref{eq7})  can be used to
 obtain the solution to the  system of equations (\ref{eq1}) and
(\ref{eq2}) everywhere in the $(x,t)$ plane except  in the regions
corresponding to  {\it simple wave} solutions where $v=v(c)$
\cite{lan1}. In this case, the Jacobian
$\Delta=\partial(v,w)/\partial(x,t)$ vanishes identically. In
deriving  Eq.\ (\ref{eq10}), we divided Eq.\ (\ref{eq8}) by this
Jacobian, and the solution for which $\Delta=0$ is not recovered.
Thus, a simple wave solution cannot be recovered from the general
equation (\ref{eq12}).

 If $v$ is a function of
$\lambda$ only, as in a simple wave, we can rewrite Eqs.\
(\ref{eq1}) and (\ref{eq2}) as \cite{lan1}
\begin{eqnarray}
&&{\partial \lambda \over \partial t}+{ d (\lambda v) \over d \lambda}{\partial \lambda\over \partial x}=0,\label{eq21} \\
&&{\partial v \over \partial t}+\left( v+{dw\over
dv}\right){\partial v\over \partial x}=0.\label{eq22}
\end{eqnarray}
Since
\begin{eqnarray}
{\partial \lambda /\partial t \over \partial \lambda /\partial x
}=-\left({\partial x\over\partial t}\right)_\lambda,\ \ \ \
{\partial v /\partial t \over \partial v /\partial x
}=-\left({\partial x\over\partial t}\right)_v, \nonumber
\end{eqnarray}
we obtain from Eqs.\ (\ref{eq21}) and (\ref{eq22})
\begin{eqnarray}
&&\left({\partial x\over\partial t}\right)_\lambda=v+\lambda{d v\over d\lambda},\label{eq23} \\
&&\left({\partial x\over\partial t}\right)_v=v+{d w\over d v
}.\label{eq24}
\end{eqnarray}
However, because $v=v(\lambda)$, it follows that $(\partial x
/\partial t)_\lambda=(\partial x /\partial t)_v$, so that $\lambda
dv/d\lambda=dw/dv=(dw/d\lambda)(d\lambda/dv)=(c^2/\lambda)d\lambda/dv$,
and therefore
\begin{equation}
v=\pm \int{c\over\lambda}d\lambda. \label{eq25}
\end{equation}
 Next, we combine Eqs. (\ref{eq23}),
(\ref{eq24}) and (\ref{eq25}) to give $(\partial x/\partial
t)_v=(\partial x/\partial t)_\lambda=v+\lambda(d v/d\lambda)=v\pm
c(v)$. Integrating with respect to $t$ then gives \cite{lan1}
\begin{equation}
x=t[v\pm c(v)]+f(v),\label{eq28}
\end{equation}
where $f(v)$ is an arbitrary function of the velocity $v$
determined from the initial conditions, and $c(v)$ is given by
Eq.\ (\ref{eq25}).

Equations (\ref{eq25}) and (\ref{eq28}) give the general solution
for the {\it simple wave}. The two signs in Eqs.\ (\ref{eq25}) and
(\ref{eq28}) correspond to the direction of  wave propagation:
$(+)$ is for a wave propagating in the positive $x$ direction, and
$(-)$ is for a wave propagating in the negative $x$ direction.

It is also convenient to solve  Eqs.\ (\ref{eq1}) and (\ref{eq2})
using the method of {\it characteristics} \cite{lan1}. Multiplying
Eq.\ (\ref{eq1}) by $c/\lambda$, and then adding and subtracting
from
 Eq.\ (\ref{eq2}), and making use of the relation $\partial w
/\partial x=(dw/d\lambda)(\partial \lambda/\partial
x)=(c^2/\lambda)(\partial \lambda/\partial x)$, we obtain
\begin{equation}
{\partial v\over \partial t} \pm {c\over\lambda}{\partial
\lambda\over
\partial t}+(v\pm c)\left({\partial v\over \partial x} \pm {c\over\lambda}{\partial
\lambda\over
\partial x}\right)=0.\label{eq29}
\end{equation}
We now introduce the new unknown functions
\begin{equation}
J_+=v+\int {c\over\lambda}d\lambda,\ \ \ \ J_-=v-\int
{c\over\lambda}d\lambda,\label{eq30}
\end{equation}
which are called {\it Riemann invariants}. In terms of $J_+$ and
$J_-$, the equations of motion take the simple form \cite{lan1}
\begin{equation}
\left[{\partial\over \partial t}+(v+c){\partial\over \partial
x}\right]J_+=0,\ \ \ \ \ \left[{\partial\over \partial
t}+(v-c){\partial\over \partial x}\right]J_-=0.\label{eq33}
\end{equation}
The differential operators acting on $J_+$ and $J_-$ are  the
operators for differentiation along the curves $C_+$ and $C_-$ (
called {\it characteristics}) in the $(x,t)$ plane given by the
equations
\begin{equation}
C_+:\ \ \ {dx\over dt}=v+c, \ \ \ \ \ C_-:\ \ {dx\over dt}=v-c. \
\ \  \label{eq34}
\end{equation}
The values of $v$ and $c$ at every point of the $(x,t)$ plane are
given by the values of the Riemann invariants  $J_+$ and $J_-$
which are transported to this point along the $C_+$ and $C_-$
characteristics from the region where the values of $J_+$ and
$J_-$ (and therefore $v$ and $c$) are known. Equations\
(\ref{eq30}) and (\ref{eq34}) are very convenient for numerical
solution of the equations of motion and also for general analysis
of the flow.

  The totality of  available space-time generally consists of the
regions where either a simple wave solution [Eq.\ (\ref{eq28})] or
a general solution [solution to Eq.\ (\ref{eq13}) together with
Eq.\ (\ref{eq7})] is applicable. The boundary between the simple
wave solution and the general solution, like any boundary between
two analytically different solutions, is a characteristic
\cite{lan1}.  In solving  particular problems (see Sec. V), the
value of the function $\chi(v,w)$ on this boundary characteristic
must be determined. The matching condition at the boundary between
the simple wave solution and the general solution is obtained by
substituting Eq.\ (\ref{eq7}) for $x$ and $t$ into the equation
for the simple wave [Eq.\ (\ref{eq28})]. This gives
\begin{equation}
{\partial \chi \over \partial v}\pm c{\partial \chi \over \partial
w}+f(v)=0.\label{eq35}
\end{equation}
Moreover, in the simple wave solution (and therefore on the
boundary characteristic), we obtain
$dw/dv=(dw/d\lambda)(d\lambda/dv)=(c^2/\lambda)(d\lambda/dv)=\pm
c$. Substituting  into Eq.\ (\ref{eq35}) then gives
\begin{equation}
{\partial \chi \over \partial v}+ {dw\over dv}{\partial \chi \over
\partial w}+f(v)={d\chi \over dv}+f(v)=0. \label{eq36}
\end{equation}
Equation\ (\ref{eq36}) can be integrated to give \cite{lan1}
\begin{equation}
\chi = - \int f(v)dv, \label{eq31}
\end{equation}
which determines the required boundary value of $\chi$.
\section{General Solution}
Here we  consider two separate cases. For the case of a cold beam
($p_0=0$ and $c_p^2=0$) it follows that  $w=c_g^2
(\lambda/\lambda_0)$ [Eq.\ (\ref{eq2a})], and $c^2(w)=\lambda
dw/d\lambda=w$. In the opposite limit where we can neglect the
electric field compared
 to the  thermal pressure, it follows that $w=(c_p^2/2)(\lambda/\lambda_0)^2$, and
$c^2(w)=\lambda dw/d\lambda=2w$. Introducing in place of $w$ the
variable $c=\sqrt{n w}$, where $n=1,2$, we can rewrite Eq.\
(\ref{eq13}) as
\begin{eqnarray}
&&{\partial^2 \chi^{n=1}\over \partial c^2}+{1\over c}{\partial
\chi^{n=1}\over \partial c}-4{\partial ^2 \chi^{n=1}\over\partial
v^2}=0,\ \ \ \text{for}\ n=1\ , \label{eq14} \\
&&{\partial^2 \chi^{n=2}\over \partial c^2}-{\partial ^2
\chi^{n=2}\over\partial v^2}=0,\ \ \ \ \ \text{for} \ n=2.
\label{eq15}
\end{eqnarray}
Equation\ (\ref{eq15}) is an ordinary wave equation whose general
solution is
\begin{eqnarray}
\chi^{n=2}(v,c)=f_1(c+v)+f_2(c-v),\label{eq16}
\end{eqnarray}
where $f_1$ and $f_2$ are arbitrary functions. To find the general
solution to Eq.\ (\ref{eq14}) we first  Fourier transform with
respect to the  $v$ dependence. This gives
\begin{equation}
{\partial^2 \chi^{n=1}_k\over \partial c^2}+{1\over c}{\partial
\chi^{n=1}_k\over \partial c}-4 k^2\chi^{n=1}_k=0. \label{eq17}
\end{equation}
Equation (\ref{eq17}) is Bessel's  equation of order zero which
has two Hankel functions, $H_0^{(1)}(2kc)$ and
$H_0^{(2)}(2kc)=[H_0^{(1)}(2kc)]^{*}$, as independent solutions.
Here $(^{*})$ represents complex conjugate. Using the integral
representation of the Hankel function \cite{abram1},
\begin{equation}
H_0^{(1)}(x)=-2i\int_1^\infty
{\exp^{ixt}\over\sqrt{t^2-1}}dt,\label{eq18}
\end{equation}
the general solution to Eq.\ (\ref{eq14}) can be expressed as
\begin{eqnarray}
\chi^{n=1}(v,c)=\int_{-\infty} ^\infty dk\int_1^\infty dt A(k)
{\exp^{i(2ct-v)k}\over\sqrt{t^2-1}}+\int_{-\infty} ^\infty
dk\int_1^\infty dt B(k)
{\exp^{i(2ct+v)k}\over\sqrt{t^2-1}},\label{eq19}
\end{eqnarray}
where $A(k)$ and $B(k)$ are arbitrary functions. Finally, we can
rewrite Eq.\ (\ref{eq19}) as
\begin{eqnarray}
\chi^{n=1}(v,c)=\int_1^\infty
{dt\over\sqrt{t^2-1}}[f_1(tc-v/2)+f_2(tc+v/2)],\label{eq20}
\end{eqnarray}
where $f_1$ and $f_2$ are arbitrary functions such that the
integrals in Eq.\ (\ref{eq20}) converge. Equation (\ref{eq20})
provides the general solution to Eq.\ (\ref{eq14}).

In the regions pertaining to the simple wave solution, Eq.\
(\ref{eq25}) gives the general relation between velocity and the
density or sound speed in the wave. In the two cases considered
here, $\lambda/\lambda_0=(c/c_g)^2$ for a cold beam $(n=2)$, and
$\lambda /\lambda_0=c/c_p$ for a pressure-dominated beam $(n=1)$.
For these two cases
\begin{eqnarray}
&&v=\pm c+a,\ \ \ \ \ \ n=1, \label{eq26} \\ &&v=\pm 2c+a,\ \ \ \
\ n=2,\label{eq27}
\end{eqnarray}
where $a$ is a constant. Equations (\ref{eq26}) and (\ref{eq27})
together with Eq.\ (\ref{eq28}) give a simple wave solution for
the two cases considered in this section. From Eq.\ (\ref{eq30}),
the corresponding Riemann invariants can be expressed as
\begin{eqnarray}
&&J_+^{n=1}=v+c,\ \ \ \ \ J_-^{n=1}=v-c,\ \ \ \ \ (n=1),
\label{eq31a}\\
&&J_+^{n=2}=v+2c,\ \ \ \ \ J_-^{n=2}=v-2c,\ \ \ \ \ (n=2).
\label{eq32}
\end{eqnarray}

\section{General Solution of the Initial Value Problem}
In this section we make use of Eqs.\ (\ref{eq16}) and (\ref{eq20})
to solve the initial value problem for the case of beam expansion
into  vacuum. The initial conditions for this problem are  zero
flow velocity at every point, $v_0(x,0)=0$, and prescribed density
profile $\lambda(x,0)=\lambda_0(x)$, which expresses the initial
line density as a function of $x$. At some later time $t=t_f$, the
density and velocity profiles will be given by  the functions
$\lambda(x,t_f)$ and $v(x,t_f)$ which are the solutions to Eqs.\
(\ref{eq1}) and (\ref{eq2}). Since the equations of motion [Eqs.\
(\ref{eq1}) and (\ref{eq2})] are time-reversible, the flow
described by $\bar \lambda(x,t)=\lambda(x,t_f-t)$ and $\bar
v(x,t)=-v(x,t_f-t)$ are also  solutions to these equations with
initial conditions $\bar v(x,0)=-v(x,t_f)$ and $\bar
\lambda(x,0)=\lambda(x,t_f)$.  At time $t=t_f$ this flow has zero
velocity profile ($\bar v=0$) and the density profile is given by
the  initial profile for the expansion problem, i.e., $\bar
\lambda(x,t_f)=\lambda_0(x)$.

\begin{figure}[htb]
\centering
\includegraphics*[width=100mm]{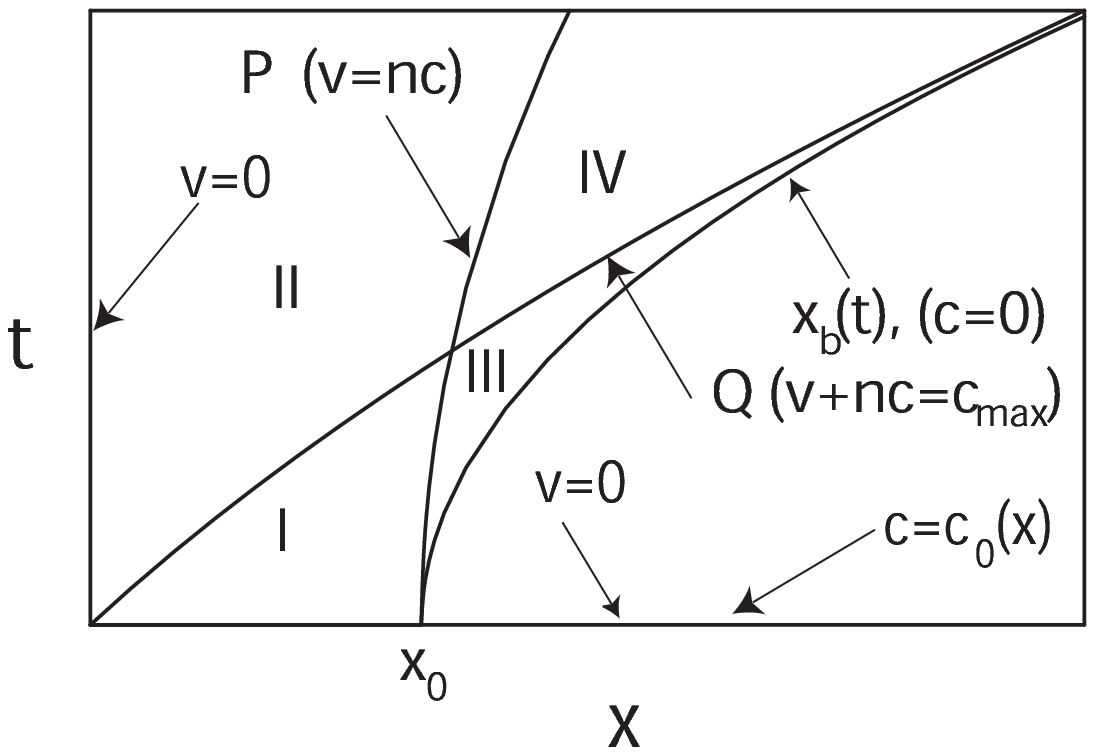}
\vspace{400pt} \caption{}\label{fig2}
\end{figure}

\begin{figure}[htb]
\centering
\includegraphics*[width=100mm]{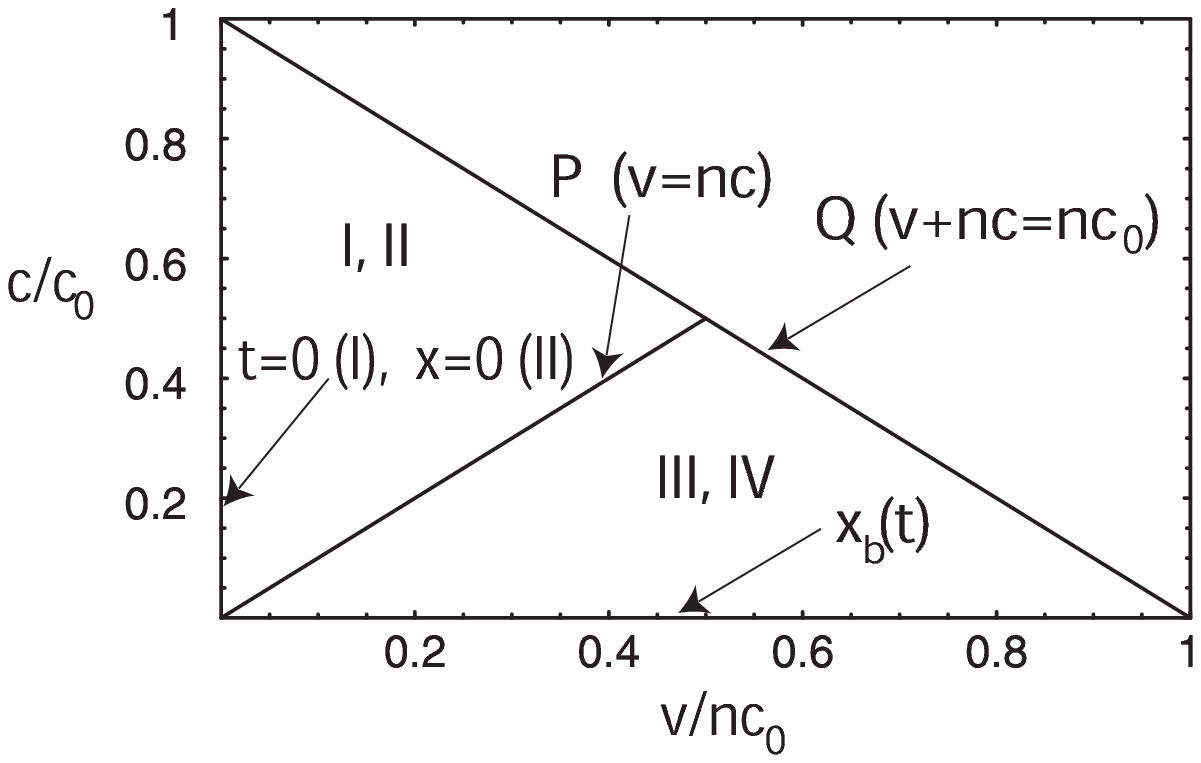}
\vspace{400pt} \caption{}\label{fig3}
\end{figure}

To solve the initial value problem we assume that the density
profile $\lambda_0(x)$, or equivalently the sound velocity profile
$c_0(x)$,  decreases monotonically to zero at the beam boundary
$x=\pm x_0$, is an even function of $x$, and is an invertable
function for $x>0$ everywhere where the density is non-zero.
Therefore, we assume that at $t=0$ the inverted profile $x_0(c)$
is known. The condition that $c_0(x)$  decreases monotonically to
zero at the beam boundary means that no  rarefaction wave is
launched from the boundary into the beam as it expands. We will
treat the case with discontinues in $c_0(x)$ at the beam boundary
in one of the examples in Sec. V. Since we are  interested in the
time-reversed problem of  beam compression, we assume that
multi-valued flow does not form as the beam expands. This is
equivalent to considering only initial density  profiles with
first derivative  decreasing continuously from the beam center to
the beam edge. This guarantees that the portions of the beam with
smaller density  accelerate faster than the portions with lower
density, and as a result, the flow is never multi-valued. We will
treat the case with multi-valued flow in one of the examples in
Sec. V.
 The region of flow in the $(x,t)$ plane and its boundaries are
illustrated in Fig.\ \ref{fig2}. It is obvious that the flow is
symmetric under reflection, $x\rightarrow -x$, and therefore we
need only to solve the equations in the region  $x>0$. In general,
there are four regions of flow. (See Figs.\ \ref{fig2} and
\ref{fig3}.) Each is separated from the others by two
characteristics, the $C_-$ characteristic $P$ on which $v=nc$, and
the $C_+$ characteristic $Q$ on which $v+nc=nc_0$. The boundary
conditions are given by
\begin{eqnarray}
v(0,t)=0,\  \ \ \ \ c(x,0)=c_0(x),\ \ v(x,0)=0,\ \text{for}\
|x|<x_0,\ \ \ c[x_b(t),t]=0, \label{eq38}
\end{eqnarray}
where $x_0=x_b(t=0)$ is the initial beam half-width, and $x_b(t)$
is the coordinate of the beam edge.

As evident from Fig. \ref{fig2}, the flow at every point in Region
I is brought to this point by the characteristics originating from
the x-axis at $t=0$. Hence, the flow in this region is fully
determined by the boundary conditions $c(x,0)=c_0(x)$ and
$v(x,0)=0$. The flow at every point in the Region II, which is
separated from Region I by the $C_+$ characteristic Q originating
from the origin in the $(x,t)$ plane, is brought to this point by
the characteristics originating from the $x=0$ line where
$v(0,t)=0$, and from Region I. Therefore, the boundary condition
for Region II is given by $v(0,t)=0$ and by the flow on the
separating characteristic Q. The flow at every point in Region
III, which is adjacent to the beam edge and separated from Region
I by the $C_-$ characteristic P originating from the beam edge at
$t=0$ in the $(x,t)$ plane, is brought to this point by the
characteristics originating from the beam edge where
$c[x_b(t),t]=0$, and from Region I. Therefore, the boundary
condition for Region III is given by $c[x_b(t),t]=0$ and by the
flow  on the separating characteristic P. The flow at every point
in Region IV, which is separated from Region II by the P
characteristic and from Region III by the Q characteristic, is
brought to this point by the characteristics originating from
Region II and Region III. Therefore the boundary condition for
Region IV is given by the flow  on the separating characteristics
P and Q.

The function $\chi(v,c)$ and Eq.\ (\ref{eq7}) provide the map of
the flow region in the $(x,t)$ plane illustrated in Fig.\
\ref{fig2}, to the $(v,c)$ plane (Fig. \ref{fig3}). The region is
a triangle $(0<c<c_0)$ limited from above by the $C_+$
characteristic Q which is  a straight line in the $(v,c)$ plane
since on this characteristic $J_+=v+nc=n c_0=const$ ($n=1,2$).
This mapping is not one-to-one. In fact Regions I and II and
Regions III and IV in the $(x,t)$ plane map into the same regions
in the $(v,c)$ plane, which means that in the $(v,c)$ plane there
will be four functions, $\chi^I$, $\chi^{II}$, $\chi^{III}$ and
$\chi^{IV}$, defined inside the area depicted in Fig.\ \ref{fig3},
which map the depicted $(v,c)$ region back into Regions I, II, III
and IV  in the $(x,t)$ plane, respectively, by means of  Eq.\
(\ref{eq7}).

Since at $t=0$, $v(x,0)=0$ and $c(0,x)=c_0(x)$, by making use of
Eq.\ (\ref{eq7}) we obtain the boundary conditions for
$\chi^I(v,c)$ in the $(v,c)$ plane in Region I, which can be
expressed as
\begin{eqnarray}
t=0=\left({\partial \chi^I\over \partial w}\right)_{v=0}&=&0,\ \text{or\ equivalently}, \ \left({\partial \chi^I\over \partial c}\right)_{v=0}=0, \label{eq39}\\
x_0(c)&=&-\left({\partial \chi^I\over \partial v}\right)_{v=0}.
\label{eq40}
\end{eqnarray}
Since  $v(0,t)=0$ at $x=0$, the boundary condition for
$\chi^{II}(v,c)$ is
\begin{eqnarray}
\left({\partial \chi^{II}\over \partial v}\right)_{v=0}=0.
\label{eq41}
\end{eqnarray}
The second boundary condition for $\chi^{II}(v,c)$ reflects
continuity of the mapping in Eq.\ (\ref{eq7}),
\begin{eqnarray}
&&\left({\partial \chi^{II}\over \partial
v}\right)_{v+nc=nc_0}=\left({\partial \chi^{I}\over \partial
v}\right)_{v+nc=nc_0}, \label{eq42}\\
&&\left({\partial \chi^{II}\over \partial
c}\right)_{v+nc=nc_0}=\left({\partial \chi^{I}\over \partial
c}\right)_{v+nc=nc_0}. \label{eq43}
\end{eqnarray}
By making use use of $c[x_b(t),t]=0$, the definition $c^2=\lambda
dw /d\lambda=nw$ and Eq.\ (\ref{eq7}), we obtain the boundary
condition for $\chi^{III}(v,c)$,
\begin{equation}
\left({\partial \chi^{III}\over \partial c}\right)_{c=0}=0.
\label{eq43a}
\end{equation}
The second boundary condition for $\chi^{III}(v,c)$ reflects
continuity of the mapping in Eq.\ (\ref{eq7}),
\begin{eqnarray}
&&\left({\partial \chi^{III}\over \partial
v}\right)_{v=nc}=\left({\partial \chi^{I}\over \partial
v}\right)_{v=nc}, \label{eq43b}\\
&&\left({\partial \chi^{III}\over \partial
c}\right)_{v=nc}=\left({\partial \chi^{I}\over \partial
c}\right)_{v=nc}. \label{eq43c}
\end{eqnarray}
Finally, the boundary conditions for $\chi^{IV}(v,c)$ reflects
continuity of the mapping in Eq.\ (\ref{eq7}),
\begin{eqnarray}
&&\left({\partial \chi^{IV}\over \partial
v}\right)_{v+nc=nc_0}=\left({\partial \chi^{III}\over \partial
v}\right)_{v+nc=nc_0}, \label{eq43d}\\
&&\left({\partial \chi^{IV}\over \partial
c}\right)_{v+nc=nc_0}=\left({\partial \chi^{III}\over \partial
c}\right)_{v+nc=nc_0}, \label{eq43e}
\end{eqnarray}
and
\begin{eqnarray}
&&\left({\partial \chi^{IV}\over \partial
v}\right)_{v=nc}=\left({\partial \chi^{II}\over \partial
v}\right)_{v=nc}, \label{eq43f}\\
&&\left({\partial \chi^{IV}\over \partial
c}\right)_{v=nc}=\left({\partial \chi^{II}\over \partial
c}\right)_{v=nc}. \label{eq43g}
\end{eqnarray}
 Next, we consider separately the two cases corresponding to a cold
beam, and a pressure-dominated beam.
\subsection{Pressure-dominated beam}
The general solution for the case of a pressure-dominated beam is
given by Eq.\ (\ref{eq16}). To satisfy the boundary condition in
Eq.\ (\ref{eq39}), we are required to choose
\begin{equation}
\chi^I=f(c-v)-f(c+v).\label{eq44}
\end{equation}
Substituting Eq.\ (\ref{eq44}) into Eq.\ (\ref{eq40}), we obtain
$f'(c)=x_0(c)/2$, and therefore
\begin{equation}
f(c)={1\over 2}\int_{c_0}^c x_0(\bar c)d\bar c. \label{eq45}
\end{equation}
Here,  we have chosen the integration constant so that $f(c_0)=0$.
Substituting Eq.\ (\ref{eq45}) into Eq.\ (\ref{eq44}) then gives
\begin{equation}
\chi^I={1\over 2}\int_{c+v}^{c-v}x_0(\bar c)d\bar c. \label{eq46}
\end{equation}
In Region II, to satisfy the boundary conditions in Eq.\
(\ref{eq41}), we are required to choose
\begin{equation}
\chi^{II}=g(c-v)+g(c+v).\label{eq47}
\end{equation}
To satisfy the boundary conditions in  Eqs.\ (\ref{eq42}) and
(\ref{eq43}) we  choose $g(c)=f(c)$. Hence, the solution in Region
II is given by
\begin{equation}
\chi^{II}={1\over 2}\left(\int_{c_0}^{c-v}x_0(\bar c)d\bar
c+\int_{c_0}^{c+v}x_0(\bar c)d\bar c \right). \label{eq48}
\end{equation}
It is readily shown that the solution  in Region III which
satisfies all of the boundary conditions in Eqs.\ (\ref{eq43a}),
(\ref{eq43b}) and (\ref{eq43c}) is given by
\begin{equation}
\chi^{III}=-{1\over 2}\left(\int_{c_0}^{v-c}x_0(\bar c)d\bar
c+\int_{c_0}^{c+v}x_0(\bar c)d\bar c \right),\label{eq48a}
\end{equation}
and the solution  in Region IV which satisfies all of the boundary
conditions in Eqs.\ (\ref{eq43d}), (\ref{eq43e}), (\ref{eq43f})
and (\ref{eq43g}) is given by
\begin{equation}
\chi^{IV}={1\over 2}\int_{v-c}^{c+v}x_0(\bar c)d\bar
c.\label{eq48b}
\end{equation}
 Finally, using Eq.\ (\ref{eq7}) and the definition
$c^2=\lambda dw /d\lambda=2w$, we obtain the solutions in Region
I,
\begin{eqnarray}
&&x-vt={1\over 2}[x_0(c-v)+x_0(c+v)], \nonumber \\
&&t={1\over 2c}[x_0(c-v)-x_0(c+v)],\label{eq49}
\end{eqnarray}
in Region II,
\begin{eqnarray}
&&x-vt={1\over 2}[x_0(c-v)-x_0(c+v)], \nonumber \\
&&t={1\over 2c}[x_0(c-v)+x_0(c+v)].\label{eq50}
\end{eqnarray}
in Region III,
\begin{eqnarray}
&&x-vt={1\over 2}[x_0(v-c)+x_0(c+v)], \nonumber \\
&&t={1\over 2c}[x_0(v-c)-x_0(c+v)],\label{eq51}
\end{eqnarray}
and in Region IV,
\begin{eqnarray}
&&x-vt={1\over 2}[x_0(v-c)-x_0(c+v)], \nonumber \\
&&t={1\over 2c}[x_0(v-c)+x_0(c+v)].\label{eq52}
\end{eqnarray}
 Equations (\ref{eq49}) and (\ref{eq50}) give the
implicit solution describing the expansion of a pressure-dominated
beam. We can also obtain the formulas for the asymptotic solution
as $t\rightarrow \infty$ or $c\rightarrow 0$. Indeed, for
$t\rightarrow \infty$ or $c\rightarrow 0$ the flow is almost
entirely in Region IV. Using Eq.\ (\ref{eq52}), we obtain
\begin{eqnarray}
&&t={1\over c}{\partial \chi^{IV}\over \partial c}={x_0(v)\over
c}+{x''(v)c\over 2}+O(c^3),\nonumber\\
&&x-vt=-{\partial \chi^{IV}\over \partial
v}=-x'_0(v)c+O(c^3).\label{eq52a}
\end{eqnarray}
Finally, in the leading approximation, we can rewrite Eq.\
(\ref{eq52a}) as
\begin{eqnarray}
{\lambda(x,t)\over\lambda_0}={1\over c_0 t}x_0\left({x\over t}
\right),\ \ \ \ v(x,t)={x\over t},\ \ \text{for}\  \ t\rightarrow
\infty. \label{eq52c}
\end{eqnarray}
Evidently, the density profile given by Eq.\ (\ref{eq52c}) is
correctly normalized.

\begin{figure}[htb]
\centering
\includegraphics*[width=100mm]{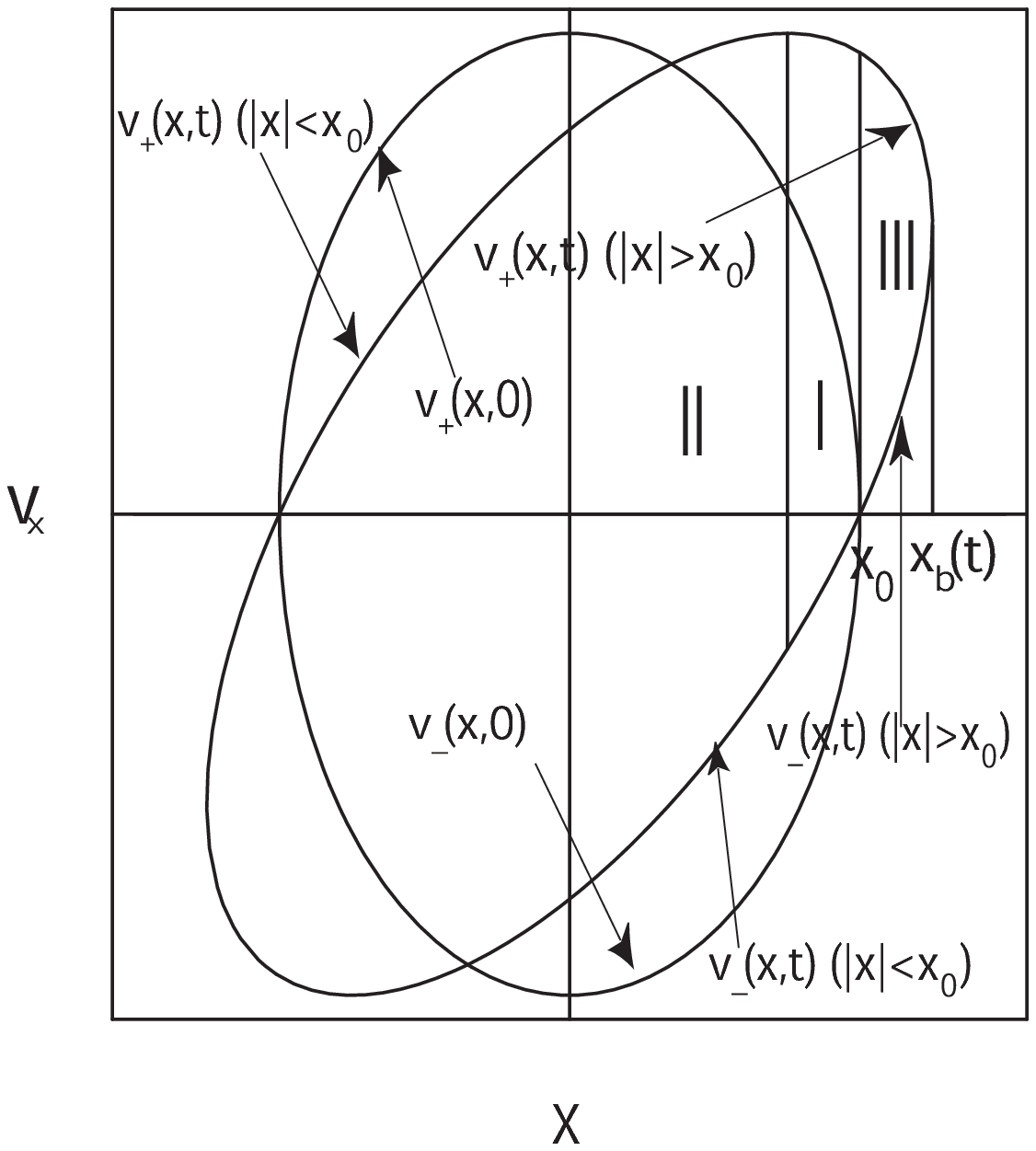}
\vspace{400pt} \caption{}\label{fig4}
\end{figure}

 The same solution can be also obtained from a
kinetic description. It's been shown in Ref. \cite{dav3} that
Eqs.\ (\ref{eq1}) and (\ref{eq2}) [together with the adiabatic
pressure relation $p=p_0(\lambda/\lambda_0)^3$] are the two key
moments of the kinetic Vlasov equation for a waterbag distribution
function ( $f=const.$ in an enclosed  area of phase space). Indeed
if we denote the upper curve in Fig.\ \ref{fig4} as $v_+(x,t)$ and
the lower curve as $v_-(x,t)$, than by multiplying the Vlasov
equation for $f(x,v_x,t)$
\begin{equation}
{\partial f\over\partial t}+v_x{\partial f\over\partial
x}=0,\label{eq53}
\end{equation}
by $1$ and by $v_x$,  integrating over $v_x$, and keeping in mind
that $f=const$ inside the region limited from above by $v_+(x,t)$
and from below by $v_-(x,t)$, we obtain
\begin{eqnarray}
&&{\partial \over \partial t}[v_+(x,t)-v_-(x,t)]+{1\over
2}{\partial
\over \partial x}[v_+(x,t)^2-v_-(x,t)^2]=0,\label{eq54}\\
&&{1\over 2}{\partial \over \partial
t}[v_+(x,t)^2-v_-(x,t)^2]+{1\over 3}{\partial \over \partial
x}[v_+(x,t)^3-v_-(x,t)^3]=0.\label{eq55}
\end{eqnarray}
Introducing the line density and flow velocity defined by
\begin{eqnarray}
&&\lambda(x,t)={\lambda_0\over [v_+(0,0)-v_-(0,0)]}[v_+(x,t)-v_-(x,t)],\label{eq56}\\
&&v(x,t)={1\over 2}[v_+(x,t)-v_-(x,t)],\label{eq57}
\end{eqnarray}
where $\lambda_0$ is the density at $x=0$ at $t=0$, we can rewrite
the Eqs.\ (\ref{eq54}) and (\ref{eq55}) in familiar form
\begin{eqnarray}
&&{\partial \over \partial t}\lambda(x,t)+{\partial\over \partial x}[\lambda(x,t)v(x,t)]=0,\label{eq58}\\
&&{\partial\over \partial t}[\lambda(x,t)v(x,t)]+{\partial \over
\partial x}[\lambda(x,t)v(x,t)^2]+{1\over 12}{ [v_+(0,0)-v_-(0,0)]^2\over \lambda_0^2}{\partial \over
\partial x}\lambda(x,t)^3=0.\label{eq59}
\end{eqnarray}
Comparing with Eq.\ (\ref{eq2}), we obtain $c^2=\lambda
dw/d\lambda=([v_+(0,0)-v_-(0,0)]^2/ 4\lambda_0^2)\lambda^2$, or
$c(x,t)=([v_+(0,0)-v_-(0,0)]/
2\lambda_0)\lambda(x,t)=(1/2)[v_+(x,t)-v_-(x,t)]$. Therefore,
$v_+(x,t)=c(x,t)+v(x,t)$ and $v_-(x,t)=v(x,t)-c(x,t)$. If the
initial profiles are given by $v(x,0)=0$ and $c(x,0)=c_0(x)$, then
$v_+(x,0)=c_0(x)$ and $v_-(x,0)=-c_0(x)$. Since Eq.\ (\ref{eq53})
represents the free-streaming motion of the particles in  phase
space along  straight-line trajectories, we readily obtain the
expressions  for $v_-(x,t)$ and $v_+(x,t)$,
\begin{eqnarray}
&&v_+(x,t)=c_0[x-v_+(x,t)t],\label{eq60}\\
&&v_-(x,t)=\pm c_0[x-v_-(x,t)t].\label{eq61}
\end{eqnarray}
Here, the $-$ sign in Eq.\ (\ref{eq61}) holds for $|x|<x_0$ ($x_0$
is the coordinate of the beam edge at $t=0$) and corresponds to
Regions I and II in Fig.\ \ref{fig2}, and the $+$ sign holds for
$x_b(t)>|x|>x_0$ and corresponds to Regions III and IV in Fig.\
\ref{fig2} (see Fig.\ \ref{fig4}). Equations (\ref{eq60}) and
(\ref{eq61}) can be rewritten as equations for $c(x,t)$ and
$v(x,t)$,
\begin{eqnarray}
&&c(x,t)={1\over 2}\left\{c_0[x-(v+c)t]\pm c_0[x-(v-c)t]\right\},\label{eq62}\\
&&v(x,t)={1\over 2}\left\{c_0[x-(v+c)t]\mp
c_0[x-(v-c)t]\right\}.\label{eq63}
\end{eqnarray}
By adding and subtracting Eqs.\ (\ref{eq62}) and (\ref{eq63}), and
inverting the resulting equations, we obtain
\begin{eqnarray}
&&\pm x_0(v+c)=x-(v+c)t,\label{eq64}\\
&&x_0(c-v)=x-(v-c)t,\label{eq65}
\end{eqnarray}
for Regions I and II, and
\begin{eqnarray}
&&\pm x_0(v+c)=x-(v+c)t,\label{eq66}\\
&&x_0(v-c)=x-(v-c)t,\label{eq67}
\end{eqnarray}
for Regions III and IV. The $+$ sign in Eqs.\
(\ref{eq64})-(\ref{eq67}) corresponds to Region I [Eqs.\
(\ref{eq64}) and (\ref{eq65})] and Region III [Eqs.\ (\ref{eq66})
and (\ref{eq67})],  and the $-$ sign corresponds to  Region II
[Eqs.\ (\ref{eq64}) and (\ref{eq65})] and Region IV [Eqs.\
(\ref{eq66}) and (\ref{eq67})] (see Figs.\ \ref{fig2} and
\ref{fig4}). The $\pm$ signs appear here because we have assumed
an even initial profile $c_0(-x)=c_0(x)$.

Finally, by adding and subtracting, Eqs.\
(\ref{eq64})-(\ref{eq67}) take the form shown in  Eqs.\
(\ref{eq49})-(\ref{eq52}).
\subsection{Cold beam}
Here we  use the general solution in Eq.\ (\ref{eq20}) to solve
the same initial value problem as discussed in the previous
section, applied now to the case of a cold beam. To satisfy the
boundary condition in Eq.\ (\ref{eq39}) we are required to choose
\begin{equation}
\chi^I(v,c)=\int_1^{\infty}{dt\over
\sqrt{t^2-1}}[f(tc+v/2)-f(tc-v/2)].\label{eq68}
\end{equation}
Substituting Eq.\ (\ref{eq68}) into Eq.\ (\ref{eq40}), we obtain
\begin{equation}
x_0(c)=-{1\over 2c}\left({\partial \chi^I\over\partial
c}\right)_{v=0}=-\int_{c}^{\infty}{dz\over\sqrt{z^2-c^2}}{df(z)\over
dz}.\label{eq69}
\end{equation}
Equation\ (\ref{eq69}) can be inverted by using  the integral Abel
transform in Appendix A. This gives
\begin{equation}
f(z)={2\over \pi}\int_z^{c_0}{q x_0(q)dq\over
\sqrt{q^2-z^2}}\Theta (z<c_0),\label{eq70}
\end{equation}
where $c_0=c_0(x=0)$, $\Theta(z<a)$ is the Heaviside
step-function, and $z>0$. Note from Eq.\ (\ref{eq68}) that in
Regions I and II, where $v<2c$, the argument of $f$ in Eq.\
(\ref{eq68}) is positive, and we can use the form of $f$ defined
in Eq.\ (\ref{eq70}). For $v>2c$ (Regions III and IV), the
argument of the function under the integral in the first term in
general solution in Eq.\ (\ref{eq20}) can become negative. Next,
we  show that the solution of the form in Eq.\ (\ref{eq68}) with
$f(z)$ continued to the regions where $z<0$ as $f(z)=f(-z)$, or
$f(z)=f(|z|)$, will satisfy the boundary conditions in Eqs.\
(\ref{eq43a})-(\ref{eq43c}). Indeed, by expanding $f$ in a Taylor
series in  Eq.\ (\ref{eq68}) for $c\rightarrow 0$, we obtain
\begin{equation}
I^-=\int^{(c_0+v/2)/c}_1{dt\over
\sqrt{t^2-1}}f(tc-v/2)=\left[f(-v/2)+{c^2\over 4}
f''(-v/2)\right]\ln\left({c_0+v/2\over
c}\right)+O(c^2),\label{eq70a}
\end{equation}
and therefore
\begin{equation}
\chi^{III} = f(v/2)\ln\left({c_0-v/2\over
c}\right)-f(-v/2)\ln\left({c_0+v/2\over
c}\right)+O\left[c^2\ln\left({c_0\over c
}\right)\right].\label{eq70b}
\end{equation}
Differentiating with respect to $c$, and taking the limit
$c\rightarrow 0$ in Eq.\ (\ref{eq70b}), we obtain
\begin{eqnarray}
\left({\partial \chi^{III} \over \partial c}\right)_{c\rightarrow
0} ={[f(-v/2)-f(v/2)]\over c}+O\left[c\ln\left({c_0\over c
}\right)\right]=0,\label{eq70bb}
\end{eqnarray}
provided $f=f(|z|)$. It readily follows that the continuity
conditions in Eqs.\ (\ref{eq43b})-(\ref{eq43c}) are also
satisfied. Therefore, the solutions in Regions I and III are given
by
\begin{equation}
\chi^{I,\ III}(v,c)=\int_1^{\infty}{dt\over
\sqrt{t^2-1}}[f(tc+v/2)-f(tc-v/2)].\label{eq70c}
\end{equation}
where
\begin{equation}
f(z)={2\over \pi}\int_{|z|}^{c_0}{q x_0(q)dq\over
\sqrt{q^2-z^2}}\Theta (|z|<c_0).\label{eq70d}
\end{equation}

To obtain the solutions in Regions II and IV, we note that the
function
\begin{equation}
\chi^{II,\ IV}(v,c)=-\int_1^{\infty}{dt\over
\sqrt{t^2-1}}[f(tc+v/2)+f(tc-v/2)].\label{eq70e}
\end{equation}
satisfies the condition in Eq.\ (\ref{eq41}). Also, if we choose
$f$ as in Eq.\ (\ref{eq70d}), the second term (and its first
derivatives) in both equations (\ref{eq68}) and (\ref{eq70e}) is
zero on the dividing characteristic  $2c+v=2c_0$, and therefore
all of the continuity conditions in Eqs.\
(\ref{eq43d})-(\ref{eq43g}) are also satisfied. Equations
(\ref{eq70c})-(\ref{eq70e}) together with Eq.\ (\ref{eq7}) give
the formal solution of the expansion problem for the case of a
cold beam. Finally, substituting Eq.\ (\ref{eq70d}) into Eqs.\
(\ref{eq70c}) and (\ref{eq70e}), changing the order of
integration, and performing the integrations, we obtain
\begin{eqnarray}
\chi^I(v,c)=&-&{2c\over \pi}\int_{1-v/2c}^{1+v/2c}dq\sqrt{q}x_0(c
q)K\left[{(v/2c)^2-(q-1)^2\over 4q}\right],\label{eq71} \\
\chi^{II}(v,c)=&-&{2c\over
\pi}\int_{1-v/2c}^{1+v/2c}dq\sqrt{q}x_0(c
q)K\left[{(v/2c)^2-(q-1)^2\over 4q}\right]\nonumber\\
&-&{8c\over \pi}\int_{1+v/2c}^{c_0/c}{dqqx_0(c q)\over
\sqrt{(q+1)^2-(v/2c)^2}}K\left[{(q-1)^2-(v/2c)^2\over
(q+1)^2-(v/2c)^2 }\right],\label{eq72} \\
\chi^{III}(v,c)=&-&{2c\over
\pi}\int_{v/2c-1}^{1+v/2c}dq\sqrt{q}x_0(c
q)K\left[{(v/2c)^2-(q-1)^2\over 4q}\right]\nonumber\\
&-&{4c\over \pi}\int_{0}^{v/2c-1}{dqqx_0(c q)\over
\sqrt{(v/2c)^2-(q-1)^2}}K\left[{4q\over (v/2c)^2-(q-1)^2}\right],\label{eq73} \\
\chi^{IV}(v,c)=&-&{2c\over
\pi}\int_{v/2c-1}^{1+v/2c}dq\sqrt{q}x_0(c
q)K\left[{(v/2c)^2-(q-1)^2\over 4q}\right]\nonumber\\
&-&{4c\over \pi}\int_{0}^{v/2c-1}{dqqx_0(c q)\over
\sqrt{(v/2c)^2-(q-1)^2}}K\left[{4q\over
(v/2c)^2-(q-1)^2}\right] \nonumber \\
&-&{8c\over \pi}\int_{1+v/2c}^{c_0/c}{dqqx_0(c q)\over
\sqrt{(q+1)^2-(v/2c)^2}}K\left[{(q-1)^2-(v/2c)^2\over
(q+1)^2-(v/2c)^2 }\right]. \label{eq74}
\end{eqnarray}
Here, $K$ is the complete  elliptic integral of the first kind
\cite{abram1}. In Sec. V, we illustrate the application of these
solutions with several examples.

We can also obtain the formulas for the asymptotic solution as
$t\rightarrow \infty$ or $c\rightarrow 0$. Indeed, for
$t\rightarrow \infty$ or $c\rightarrow 0$, the flow is almost
entirely in Regions II and IV. Using Eqs.\ (\ref{eq70a}) and
(\ref{eq70e}), we obtain
\begin{equation}
\chi^{II,\ IV}(v,c)=-I^-(v,c)-I^-(-v,c)=-\left[f(v/2)+{c^2\over 4}
f''(v/2)\right]\ln\left({c_0^2-(v/2)^2\over
c^2}\right)+O(c^2),\label{eq74a}
\end{equation}
and therefore
\begin{eqnarray}
&&t={1\over 2c}{\partial \chi^{II,\ IV}\over \partial
c}={f(v/2)\over c^2}-{f''(v/2)\over 4}\ln\left({c_0^2\over c^2
}\right)+O(1),\nonumber\\
&&x-vt=-{\partial \chi^{II,\ IV}\over \partial v}={f'(v/2)\over 2}
\ln\left({c_0^2\over c^2 }\right)+O(1).\label{eq74c}
\end{eqnarray}
Finally, in the leading approximation, we can rewrite Eq.\
(\ref{eq74c})  as
\begin{eqnarray}
{\lambda(x,t)\over\lambda_0}={x_0\over c_0 t}g\left({x\over 2 t
c_0} \right),\ \ \ \ v(x,t)={x\over t},\ \ \text{for}\  \
t\rightarrow \infty, \label{eq74d}
\end{eqnarray}
where $f(z)=c_0x_0 g(z/c_0)$ and
\begin{eqnarray}
g(z)={2\over \pi}\int_0^{\bar \lambda_0^{-1}(z^2)}\sqrt{\bar
\lambda_0(\bar x)-z^2}d\bar x. \label{eq74e}
\end{eqnarray}
Here $\bar \lambda_0(\bar x)=\lambda_0(x/x_0)/\lambda_0$ is the
scaled initial line density profile.  One can readily verify
 that the density profile given by Eq.\ (\ref{eq74d}) is
correctly normalized, and that $g(-z)=g(z)$.
\section{Examples with Different Initial Density Profiles}
In this section, we apply the formalism developed in Sec. II-IV to
several examples with different initial density profiles.
\subsection{Parabolic density profile}
As a first example, we  consider here the case of an initial
 parabolic density profile for $\lambda(x,0)=\lambda_0(x)$ with
\begin{equation}
{\lambda_0(x)\over\lambda_0}=\left[1-\left({x\over
x_0}\right)^2\right]\Theta(|x|<x_0).\label{eq75a}
\end{equation}
\subsubsection{Cold beam}
For a cold beam, $c^2=\lambda
dw/d\lambda=c_g^2(\lambda/\lambda_0)$. Substituting Eq.\
(\ref{eq75a}) into Eq.\ (\ref{eq74e}) and integrating, we obtain
\begin{equation}
f(z)={c_0x_0\over 2}\left[1-\left({z\over
c_0}\right)^2\right]\Theta(z<c_0).\label{eq76}
\end{equation}
Next we substitute Eq.\ (\ref{eq76}) into the integral
\begin{eqnarray}
I^-(a,b)=\int_1^{\infty}{dt\over \sqrt{t^2-1}}f(tc-v/2)={c_0
x_0\over
(a+b)^2}\int_1^b{dt\over \sqrt{t^2-1}}(b-t)(a+t)\nonumber \\
={c_0x_0 \over (a+b)^2}\left[\sqrt{b^2-1}(b-2a)+\left(2 a
b-1\right)\ln(b+\sqrt{b^2-1})\right] ,\label{eq77}
\end{eqnarray}
and define
\begin{equation}
I^+(a,b)=\int_1^{\infty}{dt\over
\sqrt{t^2-1}}f(tc+v/2)=I^-(b,a),\label{eq78}
\end{equation}
where we have introduced new variables
\begin{equation}
b={2c_0+v\over 2 c},\ \ \ a={2c_0-v \over 2 c}.\label{eq79}
\end{equation}
In term of the new variables, it follows that
\begin{eqnarray}
&&\chi^{I,\ III}(a,b)=I^+(a,b)-I^-(a,b)=\nonumber \\
&&{c_0x_0\over
(a+b)^2}\left[(a-2b)\sqrt{a^2-1}-(b-2a)\sqrt{y^2-1}+(2ab-1)\ln{a+\sqrt{a^2-1}\over
b+\sqrt{b^2-1}}\right],\label{eq80}\\
&&\chi^{II,\ IV}(a,b)=-I^+(a,b)-I^-(a,b)=\nonumber \\
&&-{c_0x_0\over
(a+b)^2}\left[(a-2b)\sqrt{a^2-1}+(b-2a)\sqrt{y^2-1}+(2ab-1)\ln{a+\sqrt{a^2-1}\over
b-\sqrt{b^2-1}}\right],\label{eq81}
\end{eqnarray}

\begin{figure}[htb]
\centering
\includegraphics*[width=100mm]{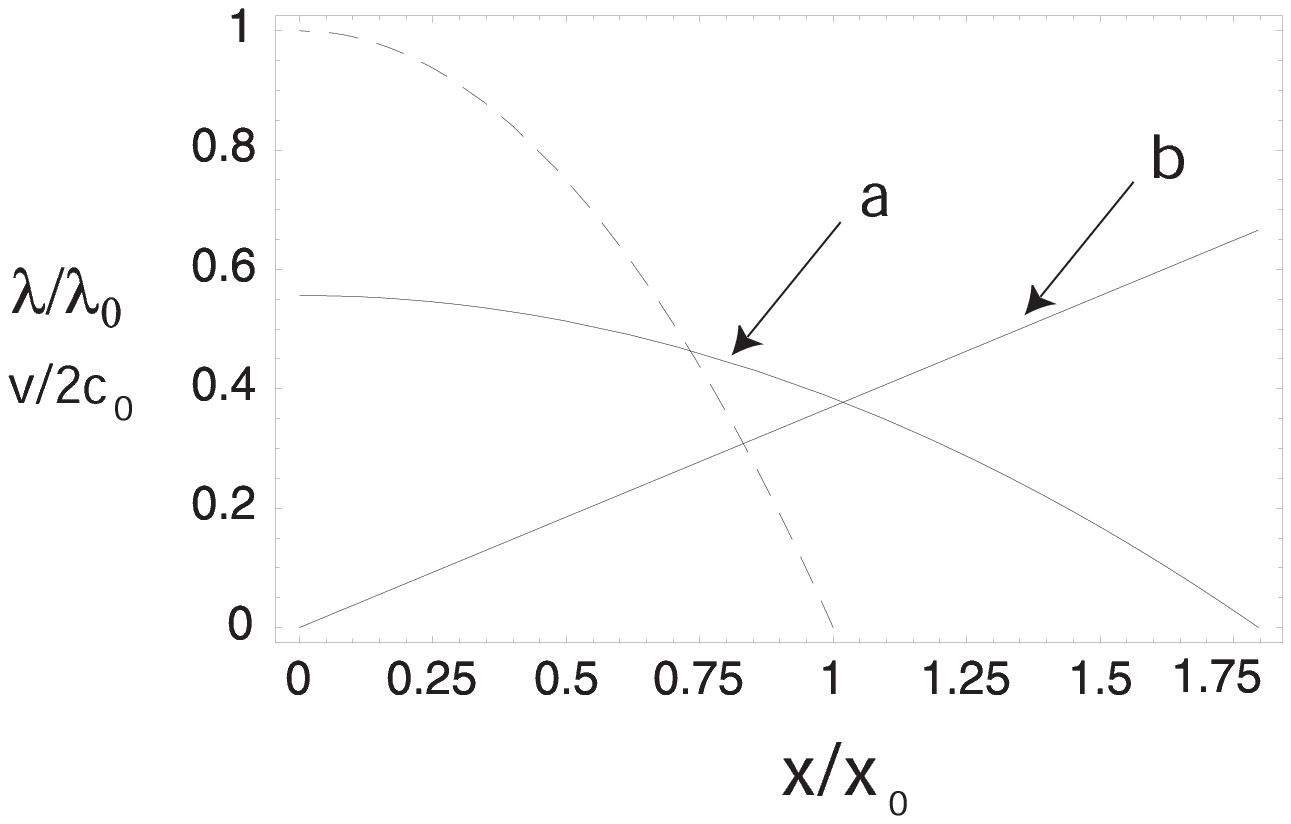}
\vspace{400pt} \caption{}\label{fig5}
\end{figure}

\begin{figure}[htb]
\centering
\includegraphics*[width=100mm]{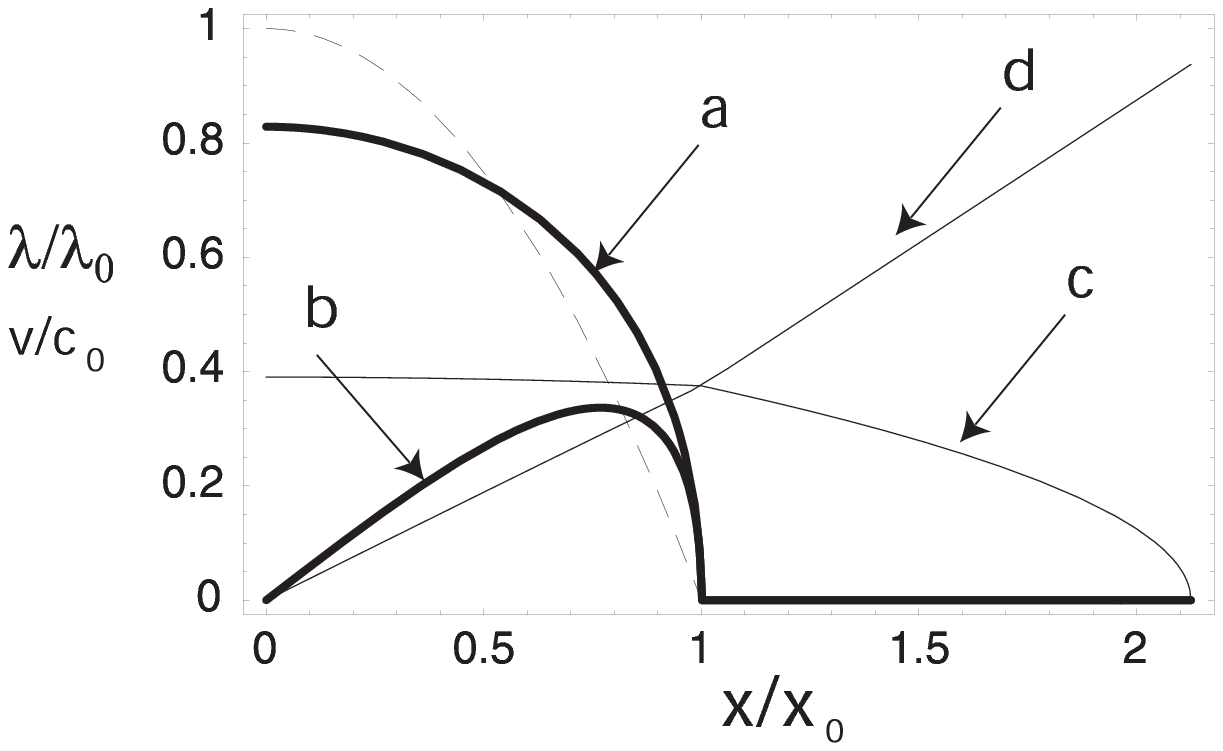}
\vspace{400pt} \caption{}\label{fig6}
\end{figure}

 By introducing the  scaled variables $\bar
x=x/x_0$, $\bar t =t c_0/x_0$ and $\bar \chi=\chi/x_0 c_0$, we can
rewrite Eq.\ (\ref{eq7}) as
\begin{eqnarray}
&&\bar t=-{(a+b)^2\over 8}\left[a{\partial \over \partial a}+b{\partial\over \partial b}\right]\bar\chi,\label{eq82}\\
&&\bar x=-{(a^2-b^2)\over 4}\left[a{\partial\over\partial
a}+b{\partial\over\partial b}\right]\bar\chi-{(a+b)\over
4}\left[{\partial\over\partial b}-{\partial\over\partial
a}\right]\bar\chi.\label{eq83}
\end{eqnarray}
 Finally, substituting
Eqs.\ (\ref{eq80}) and (\ref{eq81}) into Eqs.\ (\ref{eq82}) and
(\ref{eq83}), and using Eqs.\ (\ref{eq79}), we obtain after some
lengthy algebra the solution in Regions I and III,
\begin{eqnarray}
&\bar t&={(1-\bar v)\over 4\bar c^2}\sqrt{(1+\bar v)^2-\bar
c^2}-{(1+\bar v)\over 4\bar c^2}\sqrt{(1-\bar v)^2-\bar
c^2}+{1\over 4}\ln {1+\bar v+\sqrt{(1+\bar v)^2-\bar c^2}\over
1-\bar v+\sqrt{(1-\bar v)^2-\bar c^2}},\nonumber\\
\label{eq84}\\
&\bar x&={(\bar c^2+\bar v-\bar v^2)\sqrt{(1+\bar v)^2-\bar
c^2}+(\bar c^2-\bar v-\bar v^2)\sqrt{(1-\bar v)^2-\bar c^2}\over 2
\bar c^2},\label{eq85}
\end{eqnarray}
and in Regions II and IV,
\begin{eqnarray}
&\bar t&={(1+\bar v)\over 4\bar c^2}\sqrt{(1-\bar v)^2-\bar
c^2}+{(1-\bar v)\over 4\bar c^2}\sqrt{(1+\bar v)^2-\bar
c^2}+{1\over 4}\ln {1+\bar v+\sqrt{(1+\bar v)^2-\bar c^2}\over
1-\bar v-\sqrt{(1-\bar v)^2-\bar c^2}},\nonumber\\
\label{eq86}\\
&\bar x&={(\bar c^2+\bar v-\bar v^2)\sqrt{(1+\bar v)^2-\bar
c^2}-(\bar c^2-\bar v-\bar v^2)\sqrt{(1-\bar v)^2-\bar c^2}\over 2
\bar c^2}.\label{eq87}
\end{eqnarray}
Here we have introduced $\bar v=v/2c_0$ and $\bar c=c/c_0$.
Equations (\ref{eq84})-(\ref{eq87}) can be easily inverted. The
result is
\begin{eqnarray}
&&{v\over c_0}=2{ x \over x_0}\left({2f\over
1+f^2}\right)^2\left({1-f^2\over
1+f^2} \right),\label{eq88}\\
&&{ c\over c_0}={2f\over 1+f^2}\sqrt{1-\left[{x\over
x_0}\right]^2\left({2f\over
1+f^2}\right)^4},\label{eq89}\\
&&\left({v\over 1-f^2} \right)^2+\left({c\over f}\right)^2=\left({2c_0\over 1+f^2}\right)^2,\label{eq90}\\
&&{\lambda\over\lambda_0}=\left({2f\over
1+f^2}\right)^2\left[1-\left({x\over x_0}\right)^2\left({2f\over
1+f^2}\right)^4\right],\label{eq91}
\end{eqnarray}
where $0<f\leq 1$ is the solution of the transcendental equation
\begin{equation}
\bar t={1-f^4\over 8 f^2}-{1\over 2}\ln f.\label{eq92}
\end{equation}
The solutions in Eqs.\ (\ref{eq88})--(\ref{eq92}) describe the
familiar self-similar solution \cite{hong1,hong2,hong3} for a
parabolic density profile and is plotted in Fig.\ \ref{fig5}.
Using Eqs.\ (\ref{eq74d}) and (\ref{eq76}), we obtain  the
asymptotic solution as $t\rightarrow \infty$
\begin{eqnarray}
{\lambda(x,t)\over\lambda_0}=\bar c^2={x_0\over 2 c_0
t}\left[1-\left({x\over 2 t c_0}\right)^2 \right],\ \ \ \
v(x,t)={x\over t},\ \ \text{for}\ \ t\rightarrow \infty.
\label{eq91a}
\end{eqnarray}
The exact solution given by Eq.\ (\ref{eq91}) and asymptotic
solution given by Eq.\ (\ref{eq91a}) are compared in Fig.\
\ref{fig6} (line b) for $c_0 t/x_0=50$.

\subsubsection{Pressure-dominated beam}
\begin{figure}[htb]
\centering
\includegraphics*[width=100mm]{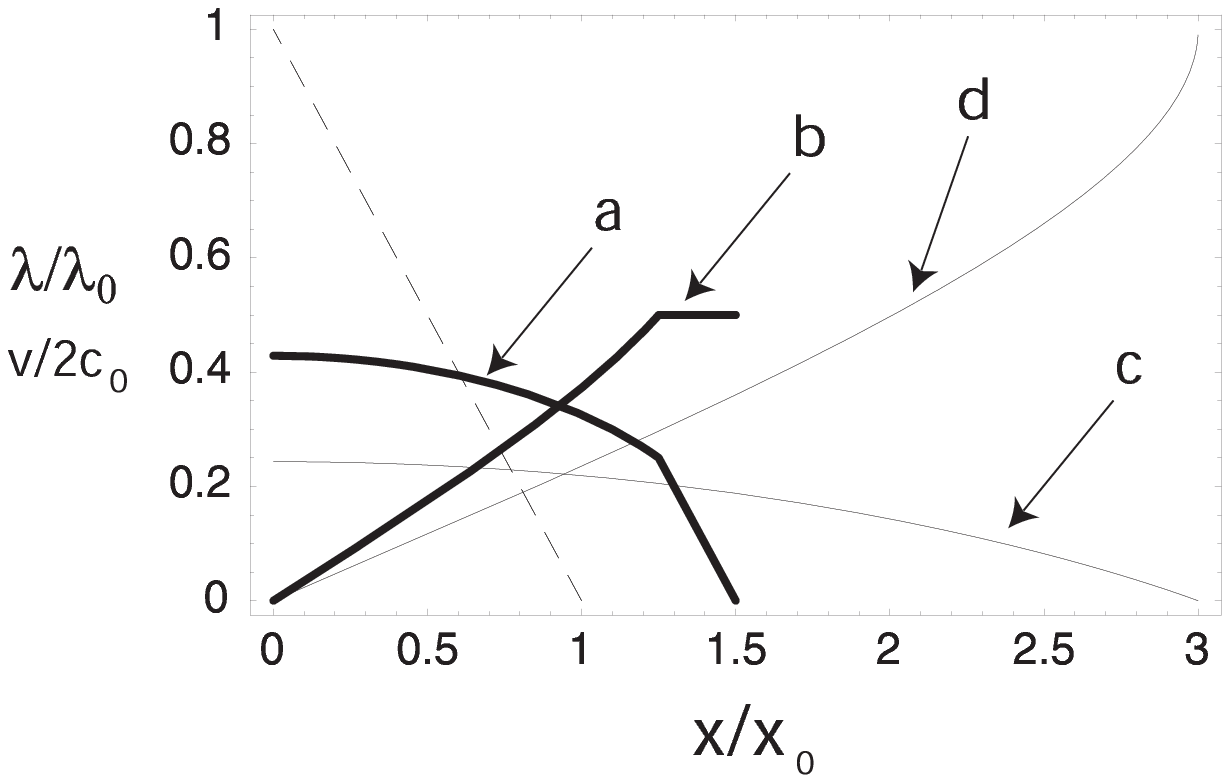}
\vspace{400pt} \caption{}\label{fig7}
\end{figure}

\begin{figure}[htb]
\centering
\includegraphics*[width=100mm]{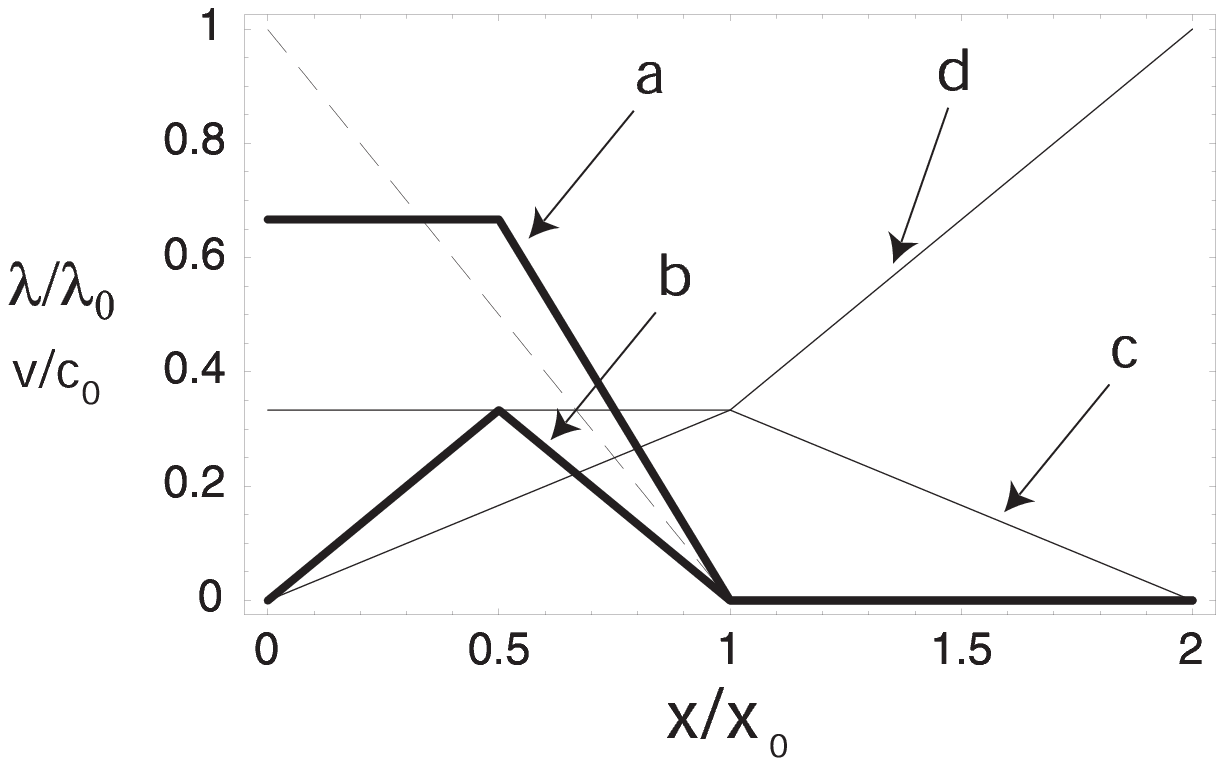}
\vspace{400pt} \caption{}\label{fig8}
\end{figure}

For a pressure-dominated  beam, $c^2=\lambda
dw/d\lambda=c_p^2(\lambda/\lambda_0)^2$ and threrefore the initial
profile for $x_0(c)$ is given by $x_0(c)/x_0=\sqrt{1-c/c_0}$.
Using Eq.\ (\ref{eq75a}) and Eqs.\ (\ref{eq62}) and (\ref{eq63})
we obtain the implicit solution in Regions I and II,
\begin{eqnarray}
&&\bar c+\bar v=1-[\bar x-(\bar v+\bar c)\bar t]^2,\label{eq142}\\
&&\bar c-\bar v=1-[\bar x-(\bar v-\bar c)\bar t]^2,\label{eq143}
\end{eqnarray}
and in Regions III and IV,
\begin{eqnarray}
&&\bar c+\bar v=1-[\bar x-(\bar v+\bar c)\bar t]^2,\label{eq144}\\
&&\bar v-\bar c=1-[\bar x-(\bar v-\bar c)\bar t]^2.\label{eq145}
\end{eqnarray}
Solving Eqs.\ (\ref{eq142})--(\ref{eq145}) for $\bar c(\bar x,\bar
t)$ and $\bar v(\bar x,\bar t)$, we obtain
\begin{eqnarray}
&&\bar c={1\over 4\bar t^2}[\sqrt{1+4\bar t^2+4\bar x \bar t}+\sqrt{1+4\bar t^2-4\bar x \bar t}-2],\label{eq146} \\
&&\bar v={\bar x \over \bar t}+{1\over 4\bar t^2}[\sqrt{1+4\bar
t^2-4\bar x \bar t}-\sqrt{1+4\bar t^2+4\bar x \bar
t}],\label{eq147}
\end{eqnarray}
for Regions I and II ($0<\bar x<1,\ 0<\bar t$), and
\begin{eqnarray}
&&\bar c={1\over 2\bar t^2}\sqrt{1+4\bar t^2-4\bar x \bar t},\label{eq148} \\
&&\bar v={\bar x \over \bar t}-{1\over 2\bar t^2},\label{eq149}
\end{eqnarray}
for Regions III and IV ($1<\bar x,\ 1/2<\bar t$).
 The solutions [Eqs.\
(\ref{eq146})--(\ref{eq149})] are illustrated in Fig.\ \ref{fig7}.
Using Eq.\ (\ref{eq52c}) we obtain  the asymptotic solution as
$t\rightarrow \infty$
\begin{eqnarray}
{\lambda(x,t)\over\lambda_0}=\bar c={x_0\over c_0
t}\sqrt{1-{x\over c_0 t} },\ \ \ \ v(x,t)={x\over t},\ \
\text{for}\ \ t\rightarrow \infty. \label{eq149a}
\end{eqnarray}
The exact solution given by Eq.\ (\ref{eq148}) and asymptotic
solution given by Eq.\ (\ref{eq149a}) are compared in Fig.\
\ref{fig8} (line b) for $c_0 t/x_0=10$.

\subsection{Linear density profile}
The next example we consider corresponds to the initial linear
density profile
\begin{equation}
{\lambda_0(x)\over\lambda_0}=\left(1-\left|{x\over
x_0}\right|\right)\Theta(|x|<x_0).\label{eq93a}
\end{equation}
\subsubsection{Cold beam}
\begin{figure}[htb]
\centering
\includegraphics*[width=100mm]{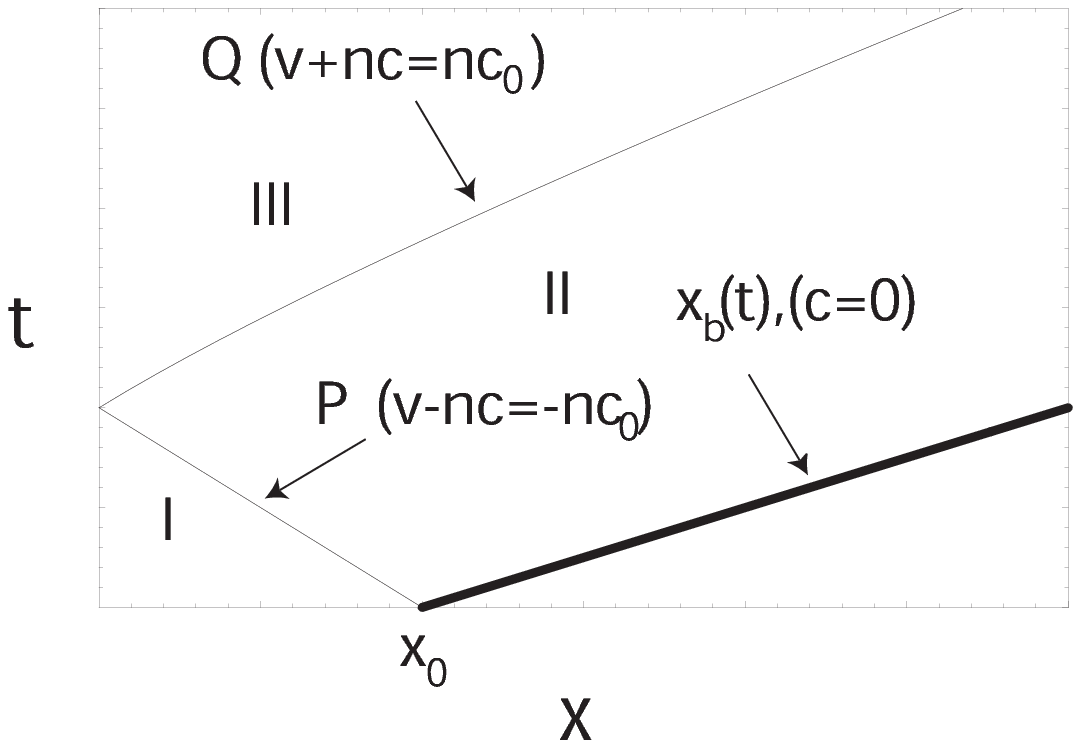}
\vspace{400pt} \caption{}\label{fig9}
\end{figure}

Here we  repeat  the intermediate steps  in the previous example.
For a cold beam, $c^2=\lambda
dw/d\lambda=c_g^2(\lambda/\lambda_0)$. Substituting Eq.\
(\ref{eq93a}) into Eq.\ (\ref{eq74e}) and integrating, we obtain
\begin{equation}
f(z)={4c_0x_0\over 3\pi}\left[1-\left({z\over
c_0}\right)^2\right]^{3/2}\Theta(z< c_0).\label{eq94}
\end{equation}
Next we substitute Eq.\ (\ref{eq94}) into the integral
\begin{eqnarray}
I^-(a,b)=\int_1^{\infty}{dt\over \sqrt{t^2-1}}f(t c-v/2)={32\over
3\pi} {c_0 x_0\over (a+b)^3}\int_1^b{dt\over
\sqrt{t^2-1}}[(b-t)(a+t)]^{3/2},\label{eq95}
\end{eqnarray}
and define
\begin{equation}
I^+(a,b)=\int_1^{\infty}{dt\over
\sqrt{t^2-1}}f(tc+v/2)=I^-(b,a),\label{eq96}
\end{equation}
where $a$ and $b$ are introduced in Eq.\ (\ref{eq79}). By
introducing the new integration variable $\alpha$ defined by
\begin{equation}
\sin(\alpha)=\sqrt{{(t-1)(b+1)\over(t+1)(b-1)}},\label{eq97}
\end{equation}
the integral in Eq.\ (\ref{eq95}) can be expressed  as
\begin{eqnarray}
{I^-(a,b)\over x_0c_0}={64\over 3\pi}\sqrt{{b-1\over
b+1}}{[(b-1)(a+1)]^{3/2}\over (a+b)^3}\int^{\pi/2}_0
d\alpha{\cos^4(\alpha)[1-k^2p^2\sin^2(\alpha)]^{3/2}\over
[1-k^2\sin^2(\alpha)]^4},\label{eq98}
\end{eqnarray}
where $k^2=(b-1)/(b+1)$ and $p^2=(a-1)/(a+1)$. The integral in
Eq.\ (\ref{eq98}) can be expressed in terms of elliptic integrals.
Finally, using the definitions $\chi^{I,\ III}=I^-(b,a)-I^-(a,b)$
and $\chi^{II,\ IV}=-I^-(a,b)-I^-(b,a)$, and using Eqs.\
(\ref{eq79}), (\ref{eq82}), and (\ref{eq83}),  we obtain after
some lengthy algebra the solution in Regions I and III,
\begin{eqnarray}
\bar t&=&{8\over \pi}{\bar v \bar c\over \sqrt{(1+\bar c)^2-\bar
v^2}}\Bigg[\Pi\left({1+\bar v -\bar c \over1+\bar v+\bar c
},{(1-\bar c)^2-\bar v^2\over(1+\bar c)^2-\bar v^2
}\right)\label{eq99}\\
&+&\Pi\left({1-\bar v -\bar c \over1-\bar v+\bar c },{(1-\bar
c)^2-\bar v^2\over(1+\bar c)^2-\bar v^2 }\right)-K\left({(1-\bar
c)^2-\bar v^2\over(1+\bar c)^2-\bar v^2 }\right)\Bigg]=2\bar v
,\nonumber
\\ \bar x&=&{4\over \pi}{(1+2\bar v^2-\bar c^2) \bar c\over
\sqrt{(1+\bar c)^2-\bar v^2}}\Bigg[\Pi\left({1+\bar v -\bar c
\over1+\bar v+\bar c },{(1-\bar c)^2-\bar v^2\over(1+\bar
c)^2-\bar v^2
}\right)\label{eq100}\\
&+&\Pi\left({1-\bar v -\bar c \over1-\bar v+\bar c },{(1-\bar
c)^2-\bar v^2\over(1+\bar c)^2-\bar v^2 }\right)-K\left({(1-\bar
c)^2-\bar v^2\over(1+\bar c)^2-\bar v^2 }\right)\Bigg]=1+2\bar
v^2-\bar c^2 ,\nonumber
\end{eqnarray}
and in Regions II and IV,
\begin{eqnarray}
\bar t&=&{4\over 3\pi}{1\over \bar c^2 \sqrt{(1+\bar c)^2-\bar
v^2}}\Bigg[6\bar c^3\bar v \Bigg\{\Pi\left({1+\bar v -\bar c
\over1+\bar v+\bar c },{(1-\bar c)^2-\bar v^2\over(1+\bar
c)^2-\bar v^2
}\right)\label{eq101}\\
&-&\Pi\left({1-\bar v -\bar c \over1-\bar v+\bar c },{(1-\bar
c)^2-\bar v^2\over(1+\bar c)^2-\bar v^2 }\right)\Bigg\}+2\bar
c(2\bar c^2+\bar v^2-3\bar c\bar v^2+\bar c-1)K\left({(1-\bar
c)^2-\bar v^2\over(1+\bar c)^2-\bar v^2 }\right)\nonumber
\\
&-&((1+\bar c)^2-\bar v^2)(2\bar c^2+\bar v^2-1)E\left({(1-\bar
c)^2-\bar v^2\over(1+\bar c)^2-\bar
v^2 }\right)\Bigg],\nonumber\\
\bar x&=&{2\over 3\pi}{1\over \bar c^2 \sqrt{(1+\bar c)^2-\bar
v^2}}\Bigg[6\bar c^3(1-\bar c^2+2\bar v^2) \Bigg\{\Pi\left({1+\bar
v -\bar c \over1+\bar v+\bar c },{(1-\bar c)^2-\bar
v^2\over(1+\bar c)^2-\bar v^2
}\right)\label{eq102}\\
&-&\Pi\left({1-\bar v -\bar c \over1-\bar v+\bar c },{(1-\bar
c)^2-\bar v^2\over(1+\bar c)^2-\bar v^2 }\right)\Bigg\}-2\bar
c\bar v(4-4\bar v^2+\bar c(2+\bar c-3\bar c^2+6\bar
v^2))K\left({(1-\bar c)^2-\bar v^2\over(1+\bar c)^2-\bar v^2
}\right)\nonumber
\\
&+&\bar v((1+\bar c)^2-\bar v^2)(4+\bar c^2-4\bar
v^2)E\left({(1-\bar c)^2-\bar v^2\over(1+\bar c)^2-\bar v^2
}\right)\Bigg],\nonumber
\end{eqnarray}
where  $\bar v=v/2c_0$, $\bar c=c/c_0$, $\bar x=x/x_0$ and $\bar t
=t c_0/x_0$. Here, $K$, $E$ and $\Pi$ are the complete elliptic
integrals of the first, second, and the third kinds, respectively
\cite{abram1}. The solution to the expansion problem for the
linear density profile Eq. (\ref{eq93a}) is given by Eqs.\
(\ref{eq99})--(\ref{eq102}) and is illustrated in Fig.\
\ref{fig9}. Using Eqs.\ (\ref{eq99}) and (\ref{eq100}), we can
rewrite the solution in Regions I and III as
\begin{eqnarray}
&&v={c_0^2\over x_0}t, \label{eq103}\\
&&{\lambda\over \lambda_0}=1+{c_0^2\over 2 x_0^2}t^2-\left|{x\over
x_0}\right|.\label{eq104}
\end{eqnarray}
This simple flow with uniform acceleration in Regions I and III
exists only until $t=t_{cr}=2x_0/c_0$. At this time the
characteristic Q overtakes the edge of the beam, and the flow for
$t>t_{cr}$ is entirely in Regions II and IV and is given by Eqs.\
(\ref{eq101}) and (\ref{eq102}). Using Eqs.\ (\ref{eq74d}) and
(\ref{eq94}) we obtain  the asymptotic solution as $t\rightarrow
\infty$
\begin{eqnarray}
{\lambda(x,t)\over\lambda_0}=\bar c^2={4 x_0\over 3\pi c_0
t}\left[1-\left({x\over 2 t c_0}\right)^2 \right]^{3/2},\ \ \ \
v(x,t)={x\over t},\ \ \text{for}\ \ t\rightarrow \infty.
\label{eq104a}
\end{eqnarray}
The exact solution given by Eqs.\ (\ref{eq101}) and (\ref{eq102})
and asymptotic solution given by Eq.\ (\ref{eq104a}) are compared
in Fig.\ \ref{fig6} (line a) for $c_0 t/x_0=50$.

\subsubsection{Pressure-dominated beam}
\begin{figure}[htb]
\centering
\includegraphics*[width=100mm]{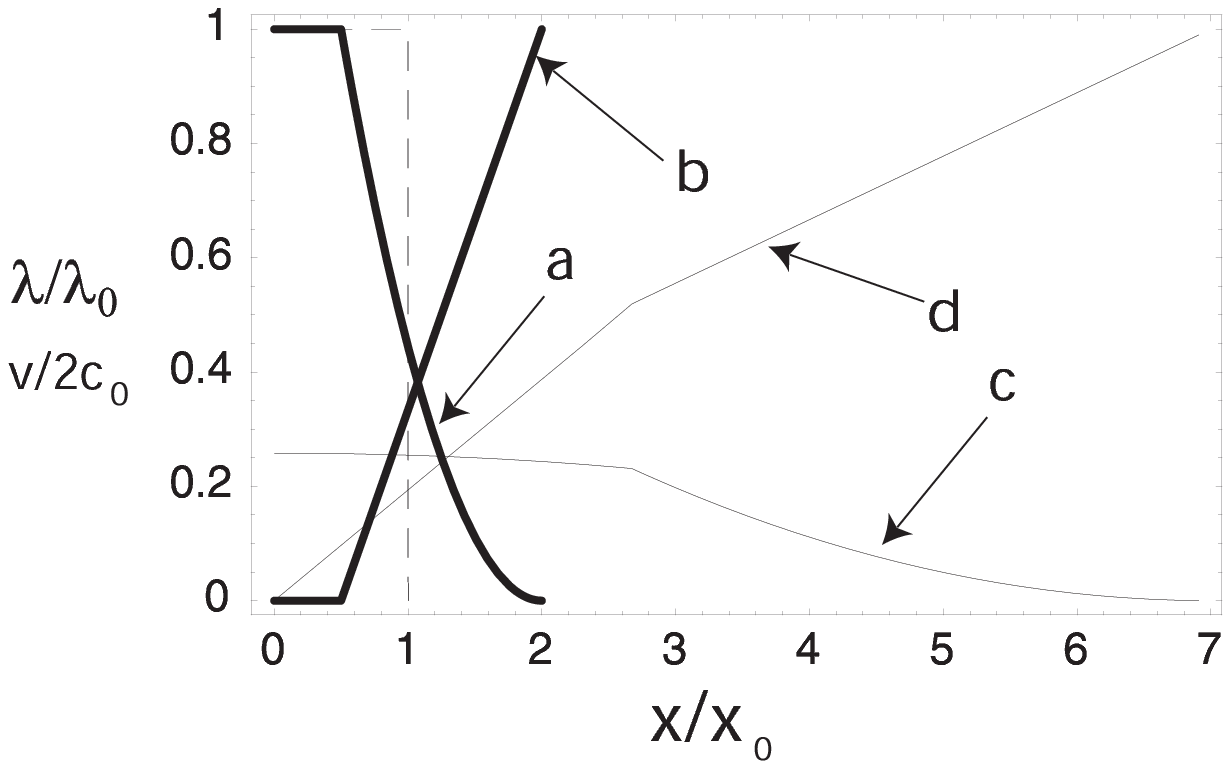}
\vspace{400pt} \caption{}\label{fig10}
\end{figure}

For a pressure-dominated  beam, $c^2=\lambda
dw/d\lambda=c_p^2(\lambda/\lambda_0)^2$. Hence, for the linear
density profile in Eq. (\ref{eq93a}, the initial profile for
$x_0(c)$ is
\begin{equation}
{x_0(c)\over x_0}=\left[1-\left({c\over
c_0}\right)\right]\Theta(c<c_0),\label{eq136}
\end{equation}
where $x_0$ is the initial beam half-length, and $c_0$ is the
sound speed at $x=0$. Substituting Eq.\ (\ref{eq136}) into Eqs.\
(\ref{eq49})--(\ref{eq52}), we obtain the solution in Region I,
\begin{eqnarray}
\bar c={1-\bar x\over 1-\bar t^2},\ \bar v=\bar t {1-\bar x\over
1-\bar t^2},\ \text{for}\ 0<\bar t<1,\ \bar t<\bar
x<1,\label{eq137}
\end{eqnarray}
in Region II,
\begin{eqnarray}
\bar c={1\over 1+\bar t},\ \bar v={\bar x\over 1+\bar t},\
\text{for}\ 0<\bar t, 0<x<\text{min}(\bar t,1), \label{eq138}
\end{eqnarray}
and in Region IV,
\begin{eqnarray}
\bar c={\bar t-\bar x\over \bar t^2-1},\ \bar v={\bar x\bar
t-1\over \bar t^2-1},\ \text{for}\ 1<\bar t, 1<x<\bar t
.\label{eq139}
\end{eqnarray}
Region III has disappeared. Indeed, using Eqs.\ (\ref{eq51}) we
obtain $\bar x=1$ and $\bar t=1$, so that the entire region
consist of just one point. The corresponding solutions [Eqs.\
(\ref{eq137})--(\ref{eq139})] are illustrated in Fig.\
\ref{fig10}. Using Eqs.\ (\ref{eq52c}) and (\ref{eq136}) we obtain
the asymptotic solution as $t\rightarrow \infty$
\begin{eqnarray}
{\lambda(x,t)\over\lambda_0}=\bar c={x_0\over c_0
t}\left(1-{x\over c_0 t} \right),\ \ \ \ v(x,t)={x\over t},\ \
\text{for}\ \ t\rightarrow \infty. \label{eq139a}
\end{eqnarray}
The exact solution given by Eq.\ (\ref{eq139}) and asymptotic
solution given by Eq.\ (\ref{eq139a}) are compared in Fig.\
\ref{fig8} (line a) for $c_0 t/x_0=10$.

\subsection{Flat-top density profile}
\begin{figure}[htb]
\centering
\includegraphics*[width=100mm]{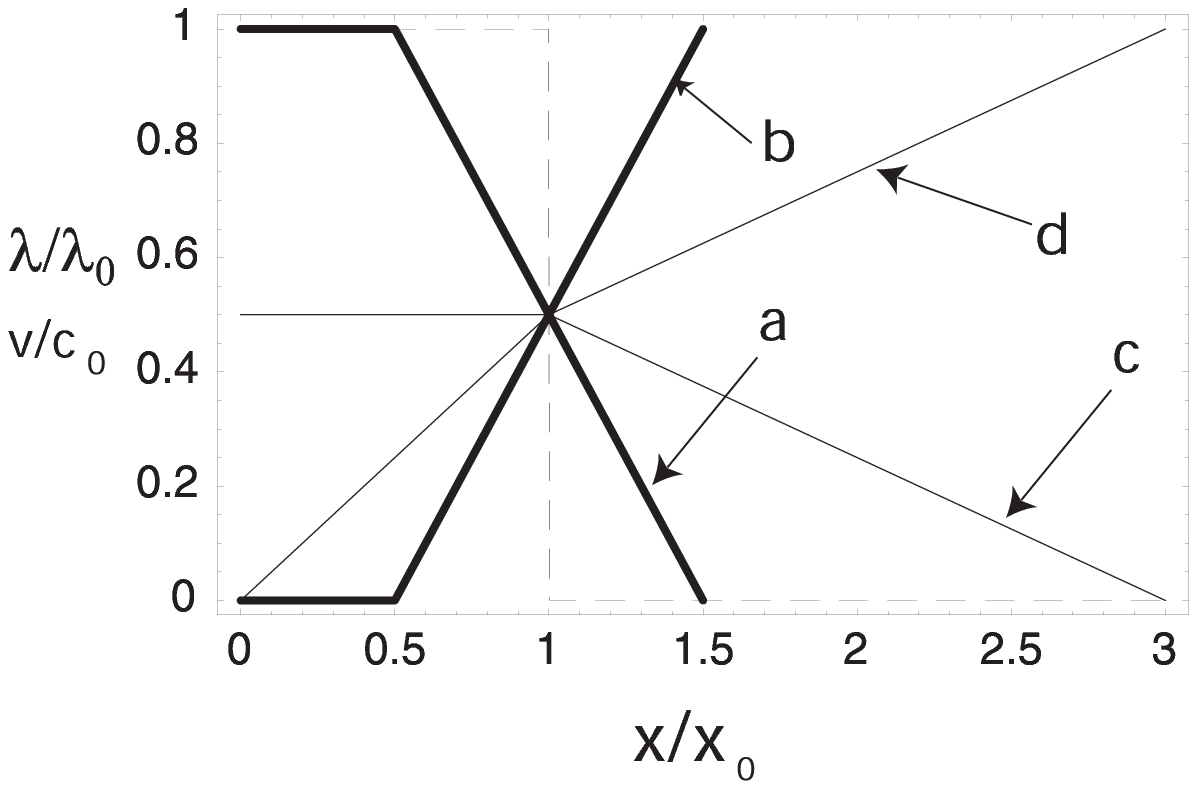}
\vspace{400pt} \caption{}\label{fig11}
\end{figure}

As discussed in Sec II, the general flow consists of regions where
either a simple wave solution [Eq.\ (\ref{eq28})] or a general
solution [solution of Eq.\ (\ref{eq13}) together with Eq.\
(\ref{eq7})] is applicable. Up to now we have considered flows
with no simple wave regions. This is guaranteed provided the
initial density profile is smooth everywhere except at the beam
edge where the sound speed is zero, $c_0(x_0)=0$. Here, we
consider an example with an initial flat-top density profile, and
a corresponding discontinuity in density at the beam edge, i.e.,
\begin{equation}
{\lambda_0(x) \over \lambda_0}=\Theta(|x|<x_0).\label{eq116}
\end{equation}
This problem is equivalent to the one-dimensional expansion of a
uniform-density gas  into vacuum in a  vessel in which the end
walls are instantly removed. The $(x,t)$ plane is shown in Fig.\
\ref{fig11}. The flow consists of three regions. In Region I the
gas is at rest, and the information that the walls have been
removed, which is carried by the $C_-$ characteristic P into the
gas, does not reach this region. Region II is the region occupied
by a simple rarefaction wave which is centered at  $t=0$ and
$x=x_0$ in the $(x,t)$ plane, and is described by the simple wave
solution in Eq.\ (\ref{eq28}). Region III is the region of
interference of this wave and and its reflection from the origin
(or another rarefaction  wave coming from the other end of the gas
region). This region is described by the general solution  in Eq.\
(\ref{eq13}) together with Eq.\ (\ref{eq7}). Regions II and III
are separated by the $C_+$ characteristic Q. On this
characteristic the boundary condition in Eq.\ (\ref{eq31}) holds.
To determine the  function $f(v)$ in Eqs.\ (\ref{eq28}) and
(\ref{eq31}), we note that for $t=0$, $x=x_0$  and therefore
$f(v)=x_0=const$. Also note that the characteristics $C_+$ bring
the value of the Rieman invariant $J_+=v+nc=nc_0=const$ to all
points of Region II. Therefore, the solution in Region II is given
by
\begin{eqnarray}
&&c={n\over n+1}c_0-{1\over n+1}{x-x_0\over t},\label{eq117}\\
&&v={n\over n+1}c_0+{n\over n+1}{x-x_0\over t},\label{eq118}
\end{eqnarray}
where $n=1,2$, and the boundary condition for $\chi$ on the
separating characteristic Q where $v+nc=nc_0$ is given by
\begin{equation}
\chi|_{v+nc=nc_0}=-x_0 v\ .\label{eq119}
\end{equation}
The second boundary condition [Eq.\ (\ref{eq41})] is given by
\begin{equation}
\left({\partial \chi\over \partial v}\right)_{v=0}=0\
.\label{eq120}
\end{equation}
\subsubsection{Cold beam}
\begin{figure}[htb]
\centering
\includegraphics*[width=100mm]{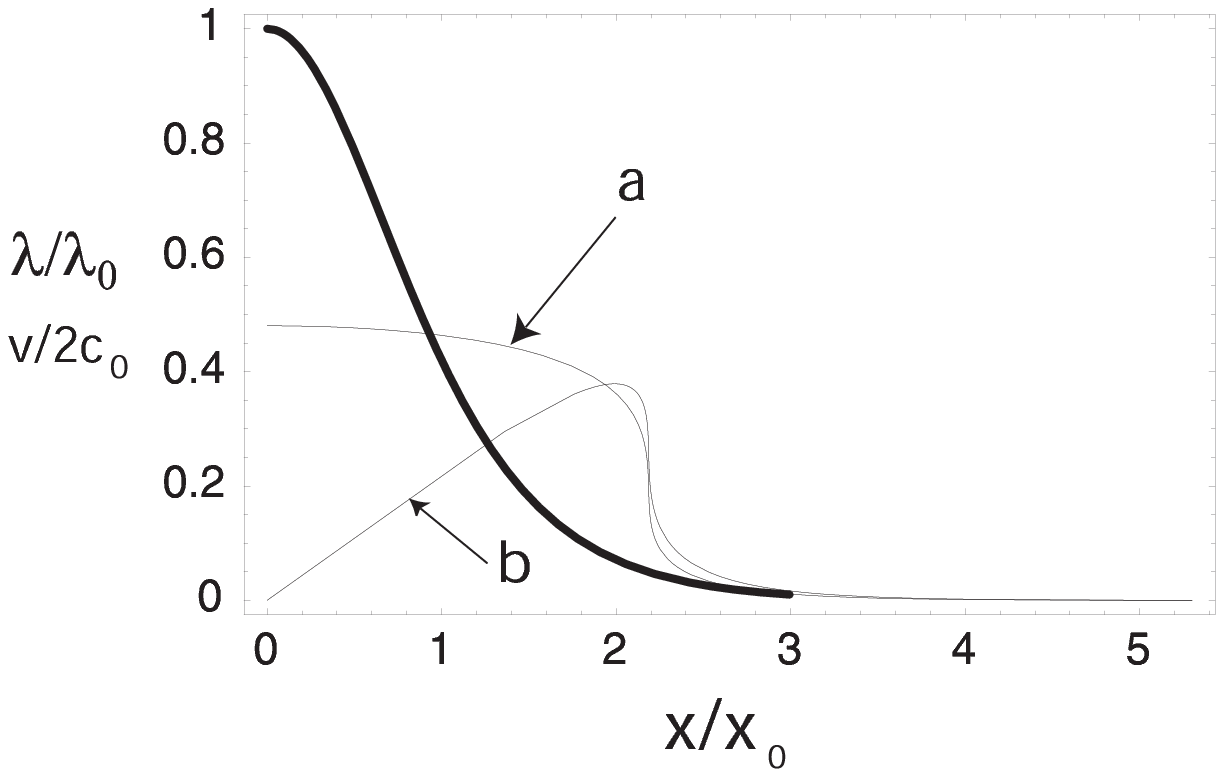}
\vspace{400pt} \caption{}\label{fig12}
\end{figure}

The function $\chi$ satisfying the boundary condition in Eq.\
(\ref{eq120}) is given by
\begin{equation}
\chi(v,c)=-\int_1^{\infty}{dt\over
\sqrt{t^2-1}}[f(tc+v/2)+f(tc-v/2)].\label{eq121}
\end{equation}
Using Eqs.\ (\ref{eq119}) and (\ref{eq121}), we obtain the
integral equation for the function $f$,
\begin{equation}
x_0 v=\int_1^{\infty}{dt\over
\sqrt{t^2-1}}\{f[(c_0-v/2)t-v/2]+f[(c_0-v/2)t+v/2]\}.\label{eq122}
\end{equation}
By setting $v=0$ in Eq.\ (\ref{eq122}) we note that the function
$f$ has the  form $f(t)=g(t)\Theta(t<c_0)$. Substituting this
expression for $f(t)$ into Eq.\ (\ref{eq122}), and changing the
integration variable to $x=t(1-v/2c_0)-v/2c_0$, we obtain the
integral equation for the function $g$,
\begin{equation}
\int^{1}_{1-v/c_0}{dx g(c_0x)\over \sqrt{(x+1)(x-[1-v/c_0])}
}=vx_0.\label{eq123}
\end{equation}
Changing the integration variables in Eq.\ (\ref{eq123}) according
to $y=\sqrt{x+1}$, and introducing the new function $p(y)\equiv
g[c_0(2y^2-1)]$, we obtain an integral equation of the Abel type,
\begin{equation}
\int^{1}_a{dy p(y)
\over\sqrt{y^2-a^2}}=c_0x_0\left(1-a^2\right)\Theta(a<1),\label{eq124}
\end{equation}
where $a=\sqrt{1-v/2c_0}$. Equation (\ref{eq124}) is easily solved
using the Abel transform described in Appendix A. We obtain
\begin{equation}
p(y)=c_0x_0{4\over \pi} y\sqrt{1-y^2}.\label{eq125}
\end{equation}
Finally, since $g(x)=p(\sqrt{1/2+x/2c_0})$, we obtain
\begin{equation}
f(x)=c_0x_0{2\over \pi}\sqrt{1-\left({x\over
c_0}\right)^2}\Theta(x<c_0).\label{eq126}
\end{equation}
Next we substitute Eq.\ (\ref{eq126}) into  the integral
\begin{eqnarray}
I^-(a,b)=\int_1^{\infty}{dt\over \sqrt{t^2-1}}f(t c-v/2)={4\over
\pi} {c_0 x_0\over (a+b)}\int_1^b{dt\over
\sqrt{t^2-1}}[(b-t)(a+t)]^{1/2},\label{eq127}
\end{eqnarray}
and define
\begin{equation}
I^+(a,b)=\int_1^{\infty}{dt\over
\sqrt{t^2-1}}f(tc+v/2)=I^-(b,a),\label{eq128}
\end{equation}
where $a$ and $b$ are introduced in Eqs.\ (\ref{eq79}). By
introducing the new integration variable $\alpha$ defined by
\begin{equation}
\sin(\alpha)=\sqrt{{(t-1)(b+1)\over(t+1)(b-1)}},\label{eq129}
\end{equation}
the integral in Eq.\ (\ref{eq127}) can be rewritten as
\begin{eqnarray}
{I^-(a,b)\over x_0c_0}={8\over \pi}\sqrt{{b-1\over
b+1}}{[(b-1)(a+1)]^{1/2}\over (a+b)}\int^{\pi/2}_0
d\alpha{\cos^2(\alpha)[1-k^2p^2\sin^2(\alpha)]^{1/2}\over
[1-k^2\sin^2(\alpha)]^2},\label{eq130}
\end{eqnarray}
where $k^2=(b-1)/(b+1)$ and $p^2=(a-1)/(a+1)$. The integral in
Eq.\ (\ref{eq130}) can be expressed in terms of elliptic
integrals. Finally, using the definition
$\chi=-I^-(b,a)-I^-(a,b)$, and making use of Eqs.\ (\ref{eq79}),
(\ref{eq82}) and (\ref{eq83}), we obtain after some lengthy
algebra the solution in Region III,
\begin{eqnarray}
&&\bar t={2\over \pi \bar c^2\sqrt{(1+\bar c)^2-\bar
v^2}}\left\{[(1+\bar c)^2)-\bar v^2]E\left[{(1-\bar c)^2-\bar
v^2\over (1+\bar c)^2-\bar v^2}\right]-2\bar c K\left[{(1-\bar
c)^2-\bar v^2\over (1+\bar c)^2-\bar
v^2}\right]\right\},\nonumber\\
\label{eq131}\\
&&\bar x={4\over \pi \bar c^2\sqrt{(1+\bar c)^2-\bar
v^2}}\Bigg\{\bar v[(1+\bar c)^2)-\bar v^2]E\left[{(1-\bar
c)^2-\bar v^2\over (1+\bar c)^2-\bar v^2}\right]-\bar v\bar
c(2+\bar c) K\left[{(1-\bar c)^2-\bar v^2\over (1+\bar c)^2-\bar
v^2}\right]\nonumber \\
&&+\Pi\left[{1-\bar c-\bar v\over 1+\bar c-\bar v},{(1-\bar
c)^2-\bar v^2\over (1+\bar c)^2-\bar v^2}\right]-\Pi\left[{1-\bar
c+\bar v\over 1+\bar c+\bar v},{(1-\bar c)^2-\bar v^2\over (1+\bar
c)^2-\bar v^2}\right]\Bigg\},\label{eq132}
\end{eqnarray}
where  $\bar v=v/2c_0$, $\bar c=c/c_0$, $\bar x=x/x_0$ and $\bar t
=t c_0/x_0$. Here, $K$, $E$ and $\Pi$ are the complete elliptic
integrals of the first, second, and the third kinds, respectively
\cite{abram1}. The solution in Region II is given by Eqs.\
(\ref{eq117}) and (\ref{eq118}) with $n=2$, i.e.,
\begin{eqnarray}
&&\bar c={1\over 3}\left(2-{\bar x-1\over \bar t}\right),\label{eq133}\\
&&\bar v={1\over 3}\left(1+{\bar x-1\over \bar
t}\right).\label{eq134}
\end{eqnarray}
Using Eqs.\ (\ref{eq133}) and (\ref{eq134}) we can determine the
trajectory of the beam edge and the characteristics Q and P. At
the beam edge, $c=0$, and therefore from Eq.\ (\ref{eq133}) we
obtain $x_b(t)=x_0+2c_0t$. On the characteristic P, $v=0$, and
from Eq.\ (\ref{eq134}) we obtain $x_P(t)=x_0-c_0t$. On the
characteristic Q, $dx/dt=v+c=4c_0/3+(x-x_0)/3t$. Integrating this
equation, we obtain
\begin{equation}
x_Q(t)=x_0+c_0t\left(2-{3\over(c_0
t/x_0)^{2/3}}\right).\label{eq135}
\end{equation}
 The solutions given by Eqs.\
(\ref{eq131})--(\ref{eq134}) are illustrated in Fig.\ \ref{fig12}.
Using Eqs.\ (\ref{eq74d}) and (\ref{eq126}) we obtain  the
asymptotic solution in Region III $[0<x<x_Q(t)]$ as $t\rightarrow
\infty$
\begin{eqnarray}
{\lambda(x,t)\over\lambda_0}=\bar c^2={2 x_0\over \pi c_0
t}\sqrt{1-\left({x\over 2 t c_0}\right)^2 },\ \ \ \ v(x,t)={x\over
t},\ \ \text{for}\ \ t\rightarrow \infty. \label{eq135a}
\end{eqnarray}
The asymptotic solution in Region II $[x_Q(t)<x<x_b(t)]$ is still
given by Eqs.\ (\ref{eq133}) and (\ref{eq134}). The exact solution
given by Eqs.\ (\ref{eq131})--(\ref{eq134}) and asymptotic
solution given by Eqs.\ (\ref{eq135a}) and (\ref{eq133}) are
compared in Fig.\ \ref{fig6} (line c) for $c_0 t/x_0=50$.
\subsubsection{Pressure-dominated beam}
\begin{figure}[htb]
\centering
\includegraphics*[width=100mm]{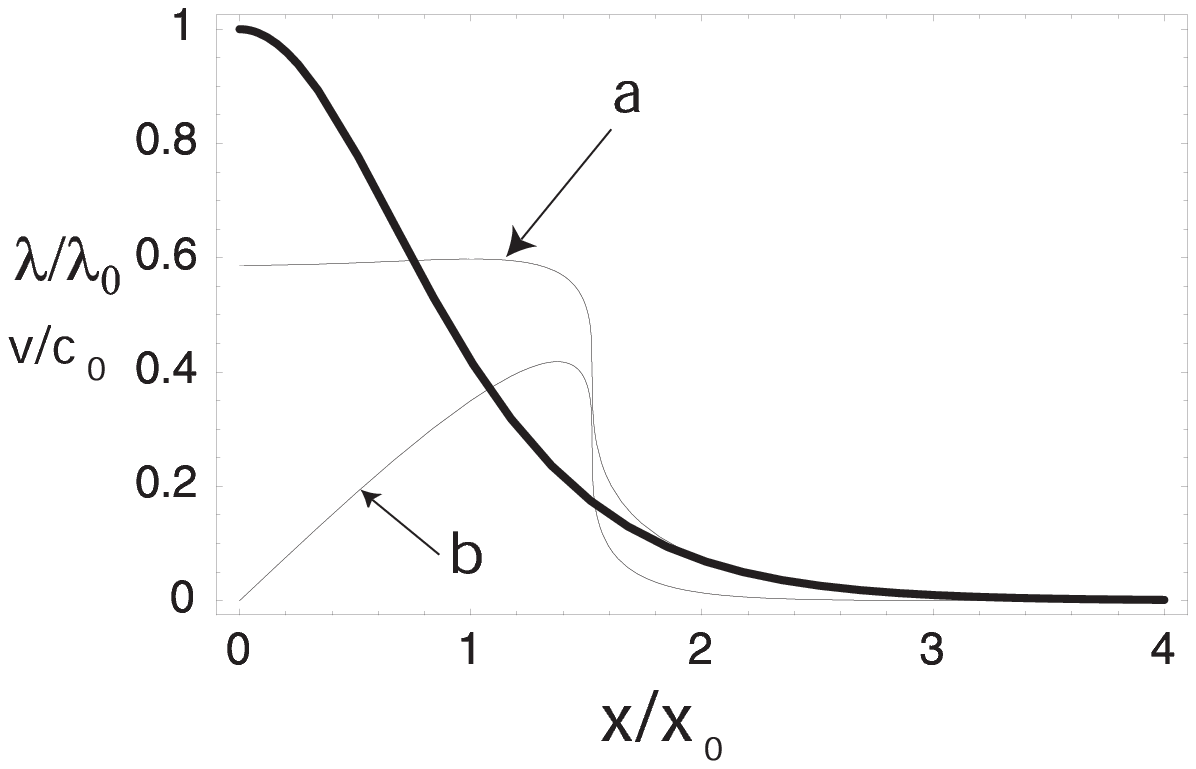}
\vspace{400pt} \caption{}\label{fig13}
\end{figure}

The function $\chi$ satisfying the boundary condition in Eq.\
(\ref{eq120}) is given by
\begin{equation}
\chi(v,c)=f(c-v)+f(c+v).\label{eq150}
\end{equation}
Using the boundary condition on the characteristic Q, it follows
that $\chi=-x_0v$ for $v+c=c_0$, and we obtain
\begin{equation}
-vx_0=f(c_0-2v)+f(c_0).\label{eq151}
\end{equation}
Substituting $v=0$ into Eq.\ (\ref{eq151}), we find that
$f(c_0)=0$ and therefore $f(c_0-2v)=-v x_0$, or
$f(x)=(x_0/2)(x-c_0)$. Using Eq.\ (\ref{eq150}), we obtain
\begin{equation}
\chi(v,c)=x_0(c-c_0),\label{eq152}
\end{equation}
and using Eq.\ (\ref{eq7}), we obtain $t=x_0/c$ and $x-vt=0$.
Therefore, the solution in Region III is given by
\begin{eqnarray}
\bar c={1\over \bar t},\ \bar v={\bar x\over \bar t},\ \ \
\text{for}\ 1<\bar t,\ 0<\bar x< \bar t-1.\label{eq153}
\end{eqnarray}
 The solution in Region II is given by Eqs.\
(\ref{eq117}) and (\ref{eq118}) with $n=1$, i.e.,
\begin{eqnarray}
\bar c={1\over 2}\left(1-{\bar x-1\over \bar t}\right),\ \bar
v={1\over 2}\left(1+{\bar x-1\over \bar t}\right),\ |\bar t
-1|<\bar x<1+\bar t. \label{eq154}
\end{eqnarray}
In Region I, the gas is undisturbed: $\bar c=1$ and $\bar v=0$,
for $0<\bar x<1-\bar t$ and $\bar t<1$. The characteristic Q is
given by the equation $\bar x_Q(t)=\bar t-1$, and the beam edge is
given by $\bar x_b(t)=1+\bar t$.
 The solutions given by Eqs.\
(\ref{eq153}) and (\ref{eq154}) are illustrated in Fig.\
\ref{fig13}. Since $\chi$ in Eq.\ (\ref{eq152}) is a linear
function of $c$ and is independent of $v$ , the asymptotic
$(t\rightarrow \infty)$ solution  coincides with the exact
solution in Eq.\ (\ref{eq153}).

\subsection{Continuous density profile (no sharp edge boundary)}
Up to this point we have considered flows which do not form shocks
and therefore are time-reversible. For such flows, the compression
problem is equivalent to the time-reversed expansion problem. In
this section, we consider an example of fluid flow which forms
shocks. For simplicity, we consider here a beam without sharp
edges in which $\lambda_0(x)$ decreases to zero monotonically  as
$x\rightarrow \pm \infty$. In particular, we consider the initial
density profile given by
\begin{equation}
{\lambda_0(x)\over\lambda_0}={1\over
\cosh^2\left(x/x_0\right)}.\label{eq105}
\end{equation}
Expanding flows  with  initially smooth profiles extending to
$x\rightarrow \pm \infty$ such as in Eq.\ (\ref{eq105}) are
entirely in Regions I and II. Indeed Regions I and II are
separated from Regions III and IV by the $C_-$ characteristic P
with $v-nc=-nc_0=0$ ($n=1,2$). However, for  profiles such as Eq.\
(\ref{eq105}), $c_0(x)>0$ for all $|x|<\infty$, and therefore the
entire region $|x|<\infty$ maps into Regions I and II in the
$(v,c)$ plane (see Fig. \ref{fig3}). As a result, the solution for
the flow is given by Eqs.\ (\ref{eq49}) and (\ref{eq50}) for
pressure-dominated beams, and by Eqs.\ (\ref{eq70c}) and
(\ref{eq70e}), with $f(z)$ defined in Eq.\ (\ref{eq70}) for  cold
beams.

\subsubsection{Cold beam}
\begin{figure}[htb]
\centering
\includegraphics*[width=100mm]{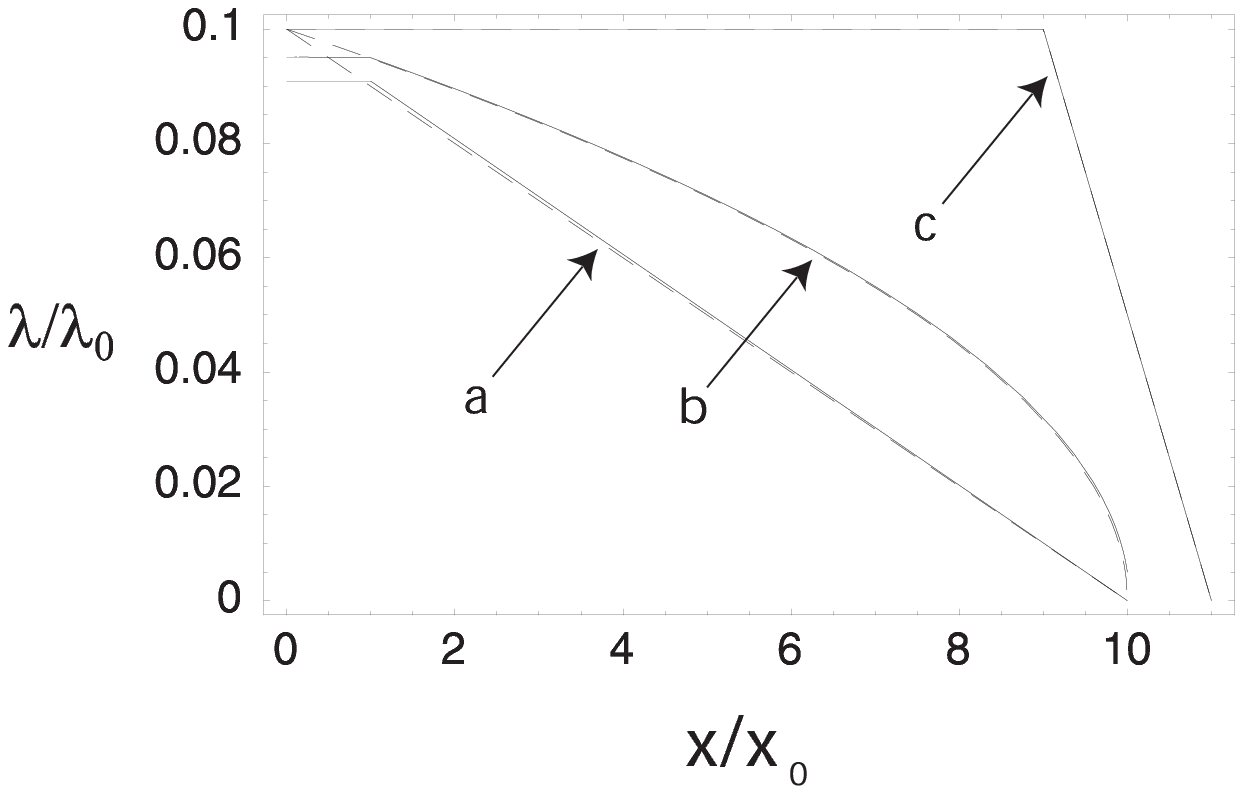}
\vspace{400pt} \caption{}\label{fig14}
\end{figure}

For a cold beam, $c^2=\lambda
dw/d\lambda=c_g^2(\lambda/\lambda_0)$, and therefore
$c/c_0=1/\cosh(x/x_0)$. Substituting Eq.\ (\ref{eq105}) into Eq.\
(\ref{eq74e}) and integrating, we obtain
\begin{equation}
f(z)=x_0 c_0\left(1-{|z|\over
c_0}\right)\Theta(z<c_0).\label{eq107}
\end{equation}
Substituting Eq.\ (\ref{eq107}) into the integral for $I^-(a,b)$
and integrating, we obtain
\begin{eqnarray}
I^-(a,b)=\int_1^{\infty}{dt\over
\sqrt{t^2-1}}f(tc-v/2)=-{2x_0c_0\over
a+b}[\sqrt{b^2-1}-b\ln(b+\sqrt{b^2-1})],\label{eq108}
\end{eqnarray}
and
\begin{equation}
I^+(a,b)=\int_1^{\infty}{dt\over
\sqrt{t^2-1}}f(tc+v/2)=I^-(b,a),\label{eq109}
\end{equation}
where $a$ and $b$ are introduced in Eq.\ (\ref{eq79}).
 Using the definitions $\chi^{I}=I^-(b,a)-I^-(a,b)$ and
$\chi^{II}=-I^-(b,a)-I^-(a,b)$, and making use of Eqs.\
(\ref{eq79}), (\ref{eq82}) and (\ref{eq83}),  we obtain the
solution in Region I,
\begin{eqnarray}
&&\bar t={\sqrt{(1+\bar v)^2-\bar c^2}-\sqrt{(1-\bar v)^2-\bar
c^2}\over 2\bar c^2},\label{eq110}\\
&&\bar x=\bar v{\sqrt{(1+\bar v)^2-\bar c^2}-\sqrt{(1-\bar
v)^2-\bar c^2}\over \bar c^2}+{1\over 2}\ln\left[{1+\bar
v+\sqrt{(1+\bar v)^2-\bar c^2}\over 1-\bar v-\sqrt{(1-\bar
v)^2-\bar c^2}}\right],\label{eq111}
\end{eqnarray}
and in Region II,
\begin{eqnarray}
&&\bar t={\sqrt{(1+\bar v)^2-\bar c^2}+\sqrt{(1-\bar v)^2-\bar
c^2}\over 2\bar c^2},\label{eq112}\\
&&\bar x=\bar v {\sqrt{(1+\bar v)^2-\bar c^2}+\sqrt{(1-\bar
v)^2-\bar c^2}\over \bar c^2}+{1\over 2}\ln\left[{1+\bar
v+\sqrt{(1+\bar v)^2-\bar c^2}\over 1-\bar v+\sqrt{(1-\bar
v)^2-\bar c^2}}\right].\label{eq113}
\end{eqnarray}
Equations (\ref{eq110})--(\ref{eq113}) can be partially inverted
to give
\begin{eqnarray}
\bar v^2={\bar t^2\bar \lambda^2(\bar t^2\bar \lambda^2+\bar
\lambda-1)\over (\bar t^2\bar \lambda^2-1)} ,\label{eq114}
\end{eqnarray}
where
\begin{eqnarray}
\bar \lambda={1\over \cosh^2(\bar x-2 \bar v \bar t)}-{\bar
v^2\over \sinh^2(\bar x-2 \bar v \bar t)},\label{eq115}
\end{eqnarray}
and  $\bar\lambda=\lambda/\lambda_0=\bar c^2$. The solutions given
by Eqs.\ (\ref{eq114}) and (\ref{eq115}) are illustrated in Fig.\
\ref{fig14}. As evident from Eq.\ (\ref{eq114}),  $(\partial \bar
v/\partial \bar \lambda)_t>0$ in some regions,  which means that
the regions with higher density accelerated faster than the
regions with lower density, and eventually multi-valued flow is
formed (see Fig.\ \ref{fig14}). This is unlike the previous
examples where $\partial \bar v/\partial \bar \lambda)_t<0$ for
all $t$, and there was no multi-valued flow.

\subsubsection{Pressure-dominated beam}
\begin{figure}[htb]
\centering
\includegraphics*[width=100mm]{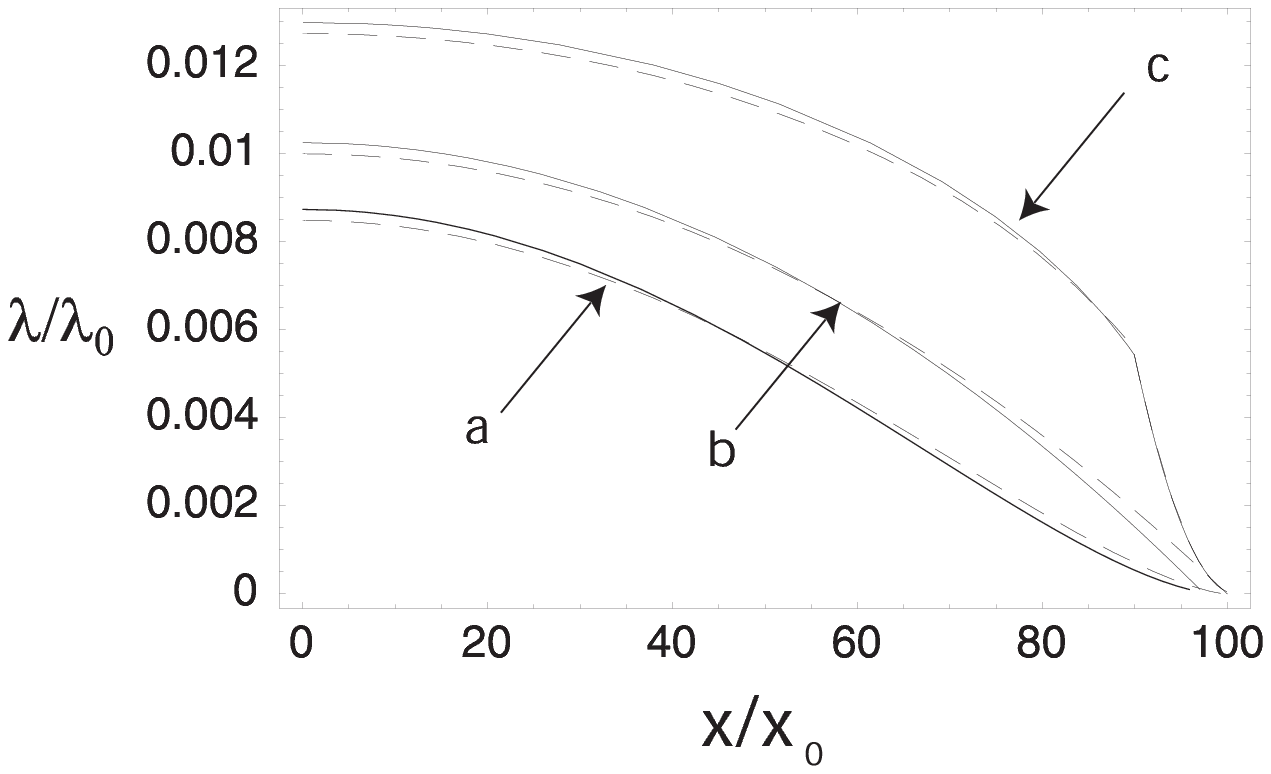}
\vspace{400pt} \caption{}\label{fig15}
\end{figure}

For a pressure-dominated  beam, $c^2=\lambda
dw/d\lambda=c_p^2(\lambda/\lambda_0)^2$. Using Eq.\ (\ref{eq105})
and Eqs.\ (\ref{eq62}) and (\ref{eq63}), we obtain the implicit
solution in Regions I and II,
\begin{eqnarray}
&&\bar c(x,t)={1\over 2}\left\{{1\over \cosh^2[\bar x-(\bar v+\bar c)\bar t]}+ {1\over \cosh^2[\bar x-(\bar v-\bar c)\bar t]}\right\},\label{eq140}\\
&&\bar v(x,t)={1\over 2}\left\{{1\over \cosh^2[\bar x-(\bar v+\bar
c)\bar t]}- {1\over \cosh^2[\bar x-(\bar v-\bar c)\bar
t]}\right\}.\label{eq141}
\end{eqnarray}
The solutions given by Eqs.\ (\ref{eq140}) and (\ref{eq141}) are
illustrated in Fig.\ \ref{fig15}.

\section{Beam shaping}
In this section we consider the beam shaping problem referred to
in Sec. I. That is, given  an initial line density profile
$\Lambda_{in}(x)$ at time $t=0$ and final line density profile
$\Lambda_{f}(x)$ at time $t=T_{shape}$, what are the initial and
finial velocity profiles, $V_{in}(x)$ and $V_f(x)$ respectively.
The beam shaping stage is necessary to prepare  the beam density
profile for the final drift compression discussed in previous
sections. Here, as in previous sections, we analyze the
time-reversed problem. Therefore, the initial density profile
$\Lambda_{in}(x)$ for the time-reversed problem is given by Eqs.\
(\ref{eq74d}) and (\ref{eq74e})  for a cold beam, or by Eq.\
(\ref{eq52c}) for a pressure-dominated beam, and the final density
profile $\Lambda_f(x)$ illustrated schematically in Fig.\
\ref{fig1}. During the beam shaping  stage, the longitudinal
pressure and electric field are negligible and the beam dynamics
is governed by  free convection decribed by
\begin{eqnarray}
&&\left({\partial \over \partial t}+v{\partial \over \partial
x}\right)v=0,\label{sh1}\\
&&{\partial x \over \partial v}-v{\partial t\over \partial
v}+\lambda {\partial t \over \partial \lambda} =0.\label{sh2}
\end{eqnarray}
Here, Eq.\ (\ref{sh2}) follows from multiplying  Eq.\ (\ref{eq11})
 by $dw/d\lambda$, and is equivalent to Eq.\ (\ref{eq1}).
Equation (\ref{sh1}) implies that the function $v$ is constant
along the characteristic given  by $dx/dt=v$, which  therefore
correspond to straight lines given by
\begin{equation}
x=vt+f(v),\  \ \ \text{or} \ \ v(x,t)=V[x-v(x,t)t],\label{sh3}
\end{equation}
where $v(x,0)=V(x)$ and $V[f(v)]=v$. Equation\ (\ref{sh3}) gives a
general solution to Eq.\ (\ref{sh1}) for the velocity profile
$v(x,t)$. Substituting Eq.\ (\ref{sh3}) into Eq.\ (\ref{sh2}),  we
obtain $t=-f'(v)+q(v)/\lambda$, where $q(v)$ is an arbitrary
function of $v$. Using the initial condition that at $t=0$,
$\lambda=\lambda_0(v)$, we obtain the solution to  Eqs.\
(\ref{sh2}), $\lambda = \lambda_0(v)/[1+t/f'(v)].$  Note that
$1/f'(v)\equiv V'[f(v)]=V'(x-vt)$, and
$\lambda_0(v)=\lambda_0[V(x-vt)]\equiv \Lambda(x-vt)$, where
$\Lambda(x)$ is the initial density profile as a function of $x$.
Therefore, the  solution to Eqs.\ (\ref{sh1}) and (\ref{sh2}) is
given by \cite{hong2}
\begin{eqnarray}
&&v(x,t)=V[x-v(x,t)t],\label{sh5}\\
&&\lambda(x,t)={\Lambda[x-v(x,t)t]\over 1+t
V'[x-v(x,t)t]}.\label{sh6}
\end{eqnarray}
Setting $t=T_{shape}$ and introducing new the function
$U_{in}(x)=x+T_{shape} V_{in}(x)$, we can rewrite Eq.\ (\ref{sh5})
as
$V_f(x)=v(x,T_{shape})=V_{in}[x-V_f(x)T_{shape}]=\left\{U_{in}[x-V_f(x)T]-[x-V_f(x)T_{shape}]\right\}/T_{shape}$,
or equivalently, $x=U_{in}[x-V_f(x)T_{shape}]$. Finally, using the
definition of the function  $U_{in}(x)$, we can rewrite Eqs.\
(\ref{sh5}) and (\ref{sh6}) in a compact and manifestly
time-reversible form giving the finial formal solution to the beam
shaping problem, i.e.,
\begin{eqnarray}
&&V_f(x)={x-U_f(x)\over T_{shape}},\ \ \ V_{in}(x)={U_{in}(x)-x\over T_{shape}},\label{sh7}\\
&&\int_0^{U_{in}(x)}\Lambda_f(\bar u)d\bar u
=\int_0^x\Lambda_{in}(\bar x)d\bar x \ \ \text{or}\
\int_0^x\Lambda_f(\bar x)d\bar x =\int_0^{U_f(x)}\Lambda_{in}(\bar
u)d\bar u.\label{sh8}
\end{eqnarray}
Here, $U_f(x)=U_{in}^{-1}(x)$ is the inverse of function
$U_{in}(x)$ such that $U_f[U_{in}(x)]\equiv x$, and we have
assumed that $\Lambda_f(-x)=\Lambda_f(x)$ and
$\Lambda_{in}(-x)=\Lambda_{in}(x)$. Examples applying results in
Eqs.\ (\ref{sh7}) and (\ref{sh8})  can be found in Ref.
\cite{hong2}.
\section{Conclusions}
To summarize, we have studied the longitudinal drift compression
of an intense charged particle beam using a one-dimensional
warm-fluid model. We have reformulated the drift compression
problem as the time-reversed expansion problem of the beam with
arbitrary line density profile and zero velocity profile. We have
obtained exact analytical solutions to the expansion problem for
the two important cases corresponding to a cold beam, and a
pressure-dominated beam, using a general formalism which reduces
the system of warm-fluid equations to a linear second-order
partial differential equation. We obtained simple approximate
analytical formulas connecting the initial and final line  density
profile and flow velocity profile for these two cases. The
asymptotic velocity profiles are linear in both cases, and
correspond to free expansion as $t\rightarrow\infty$. The scaled
density profile for a pressure-dominated beam far from the
compression point was shown to be the functional inverse of the
compressed density profile in Eq.\ (\ref{eq52c}). For a cold beam,
the profiles are connected by the Abel transform [Eqs.\
(\ref{eq74d}) and (\ref{eq74e})]. The general solution has been
illustrated for parabolic, linear, and flat-top initial
(compressed)  line density profiles. For the case of a parabolic
density profile, we have recovered the familiar self-similar
solution \cite{hong1,hong2,hong3}. We have illustrated the
formation of multi-valued flow with the exactly-solvable example
in Eq.\ (\ref{eq105}),  and identified the conditions for
shock-free compression.

 \acknowledgments

This research was supported by the U.S. Department of Energy. It
is a pleasure to acknowledge the benefit of useful discussions
with Igor Kaganovich and Hong Qin.
\appendix
\section{Abel Transform}
Here we use the following definition of the Abel transform
\begin{equation}
f(z)=A[g(z)]=2\int_z^{\infty}{g(x)xdx\over
\sqrt{x^2-z^2}}.\label{app1}
\end{equation}
The inverse Abel transform is given by
\begin{equation}
g(x)=A^{-1}[f(x)]=-{1\over \pi}\int_x^{\infty}{df(z)\over
dz}{dz\over \sqrt{z^2-x^2}}.\label{app2}
\end{equation}
The fact that Eq.\ (\ref{app2}) is indeed the inverse of the Abel
transform in Eq.\ (\ref{app1}) can be checked by direct
substitution of Eq.\ (\ref{app2}) into Eq.\ (\ref{app1}). Changing
 the order of integration leads to
\begin{eqnarray}
f(z)&=&-{1\over \pi}\int_z^{\infty}{2x
dx\over\sqrt{x^2-z^2}}\int_x^{\infty}{df(t)\over
dt}{dt\over\sqrt{t^2-x^2}}\nonumber \\
&=&-{1\over \pi}\int_z^{\infty}dt{df(t)\over dt}\int^t_z {2x
dx\over\sqrt{t^2-x^2}\sqrt{x^2-z^2}}.\label{app3}
\end{eqnarray}
Making the change of variables, $\sin
(q)=\sqrt{x^2-z^2}/\sqrt{t^2-z^2}$,  the final integral in Eq.\
(\ref{app3}) is evaluated to be
\begin{equation}
\int^t_z {2x dx\over\sqrt{t^2-x^2}\sqrt{x^2-z^2}}=\pi,\label{app4}
\end{equation}
and Eq.\ (\ref{app3}) becomes
\begin{equation}
f(z)=-\int_z^{\infty}dt{df(t)\over dt}=f(z)-f(\infty).\label{app5}
\end{equation}
Therefore, for functions $f(z)$ such that $f(\infty)=0$, Eq.\
(\ref{app5}) is an identity.

\eject

{\bf FIGURE CAPTIONS}
\newline
\newline
Fig.1:\ \ Schematic of the two stages of drift compression
described in Sec.~I, the {\it beam shaping} stage  and the {\it
drift compression} stage.
\newline
\newline
Fig.2:\ \ The area in  the $(x,t)$ plane occupied by the four
regions of flow.
\newline
\newline
Fig.3:\ \ The area in  the $(v,c)$ plane occupied by the four
regions of  flow.
\newline
\newline
Fig.4:\ \ Schematic in $(x,v_x)$ phase space of the flow for a
pressure-dominated beam.
\newline
\newline
Fig.5:\ \ Plots of the normalized density $\lambda(x,t)/\lambda_0$
(a), and the flow velocity $v/2c_0$ (b), at $c_0 t/x_0=1$  as
functions of $x$ for a cold beam. The initial density profile
(dotted line) is given by Eq.\ (\ref{eq75a}).
\newline
\newline
Fig.6:\ \ Plots of the normalized density $\lambda(x,t)/\lambda_0$
as a function of $x$ for a cold beam at $c_0 t/x_0=50$. The solid
lines are the exact solutions. The dotted lines are the
approximate solutions given by Eq.\ (\ref{eq74d}). The initial
profiles are given by: (a)
 Eq.\ (\ref{eq93a}), (b)  Eq.\ (\ref{eq75a}), and (c) Eq.\ (\ref{eq116}).
\newline
\newline
Fig.7:\ \ Plots of the normalized density $\lambda(x,t)/\lambda_0$
 for $c_0 t/x_0=0.5$ (a) and for $c_0 t/x_0=2$ (c), and the flow
velocity $v/2c_0$   for $c_0 t/x_0=0.5$ (b) and  for $c_0 t/x_0=2$
(d), as  functions of $x$ for a cold beam. The initial density
profile (dotted line) is given by Eq.\ (\ref{eq93a}).
\newline
\newline
Fig.8:\ \ Plots of the normalized density $\lambda(x,t)/\lambda_0$
as a function of $x$ for a pressure-dominated beam at $c_0
t/x_0=10$. The solid lines are the exact solutions. The dotted
lines are the approximate solutions given by Eq.\ (\ref{eq52c}).
The initial profiles are given by: (a)
 Eq.\ (\ref{eq93a}), (b) Eq.\ (\ref{eq75a}), and  (c) Eq.\ (\ref{eq116}).
\newline
\newline
Fig.9:\ \ Plots of the normalized density $\lambda(x,t)/\lambda_0$
 for $c_0 t/x_0=0.5$ (a) and  for $c_0 t/x_0=2$ (c), and the flow
velocity $v/c_0$   for $c_0 t/x_0=0.5$ (b) and  for $c_0 t/x_0=2$
(d), as functions of $x$ for a pressure-dominated beam. The
initial density profile (dotted line) is given by Eq.\
(\ref{eq75a}).
\newline
\newline
Fig.10:\ \ Plots of the normalized density
$\lambda(x,t)/\lambda_0$
 for $c_0 t/x_0=0.5$ (a) and  for $c_0 t/x_0=2$ (c), and the flow
velocity $v/c_0$   for $c_0 t/x_0=0.5$ (b) and  for $c_0 t/x_0=2$
(d), as functions of $x$ for a pressure-dominated beam. The
initial density profile (dotted line) is given by Eq.\
(\ref{eq93a}).
\newline
\newline
Fig.11:\ \ The area in  the $(x,t)$ plane occupied by the three
regions of flow.
\newline
\newline
Fig.12:\ \ Plots of the normalized density
$\lambda(x,t)/\lambda_0$
 for $c_0 t/x_0=0.5$ (a) and  for $c_0 t/x_0=3$ (c), and the flow
velocity $v/2c_0$   for $c_0 t/x_0=0.5$ (b) and for $c_0 t/x_0=3$
(d), as functions of $x$ for a cold beam. The initial density
profile (dotted line) is given by Eq.\ (\ref{eq116}).
\newline
\newline
Fig.13:\ \ Plots of the normalized density
$\lambda(x,t)/\lambda_0$ for $c_0 t/x_0=0.5$ (a) and  for $c_0
t/x_0=2$ (c), and the flow velocity $v/c_0$   for $c_0 t/x_0=0.5$
(b) and  for $c_0 t/x_0=2$ (d), as functions of $x$ for a
pressure-dominated beam. The initial density profile (dotted line)
is given by Eq.\ (\ref{eq116}).
\newline
\newline
Fig.14:\ \ Plots of the normalized density
$\lambda(x,t)/\lambda_0$ (a) and flow velocity $v/2c_0$ (b) at
$c_0 t/x_0=1.5$  as functions of $x$ for a cold beam. The initial
density profile (dotted line) is given by Eq.\ (\ref{eq105}).
\newline
\newline
Fig.15:\ \ Plots of the normalized density
$\lambda(x,t)/\lambda_0$ (a) and flow velocity $v/c_0$ (b) at $c_0
t/x_0=1.3$  as functions of $x$ for a  pressure-dominated beam.
The initial density profile (dotted line) is given by Eq.\
(\ref{eq105}).
\newline
\newline
 \eject
\thispagestyle{empty}

\eject
\thispagestyle{empty}

\eject
\thispagestyle{empty}

\eject \thispagestyle{empty}

\eject \thispagestyle{empty}

\eject \thispagestyle{empty}

\eject \thispagestyle{empty}

\eject \thispagestyle{empty}

\eject \thispagestyle{empty}

\eject \thispagestyle{empty}

\eject \thispagestyle{empty}

\eject \thispagestyle{empty}

\eject \thispagestyle{empty}

\eject \thispagestyle{empty}

 \eject
\thispagestyle{empty}

\end{document}